\documentclass[aps, prl, twocolumn, superscriptaddress]{revtex4-1} 
\usepackage{amssymb, amsmath, amsthm}
\usepackage{nicefrac}
\usepackage{xcolor}
\usepackage{graphicx}
\usepackage[protrusion=true,expansion=true]{microtype}
\usepackage{times}
\usepackage{hyperref}
\hypersetup{
    pdftitle={main},    
    pdfauthor={},
    colorlinks=true,
    citecolor=blue,
    linkcolor=blue,
    urlcolor=blue
  }
\usepackage{comment}
\usepackage{blkarray}
\usepackage{mathtools}
\usepackage{tikz}
\usetikzlibrary{decorations.pathreplacing}
\usepackage{dsfont}
\usepackage{braket}

\usepackage{graphicx}
\usepackage{dcolumn}
\usepackage{bm}

\newcommand{\prlsection}[1]{{\em {#1}.---~}}
\newcommand{\Tr}[1]{\mathrm{Tr}}

\newcommand{\e}{\text{e}}

\def\nn{\nonumber}
\newcommand{\dr}[1]{{#1}^{\rm dr}}

\usepackage{bm}
\newcommand{\be}{\begin{equation}}
\newcommand{\ee}{\end{equation}}
\newcommand{\de}{\partial}

\newcommand{\titleinfo}{
Navier-Stokes Equations for Low-Temperature One-Dimensional Fluids}
\begin{document}

\preprint{APS/123-QED}

\title{\titleinfo
}

\author{Andrew Urichuk}
\affiliation{Laboratoire de Physique Th\'eorique et Mod\'elisation, CNRS UMR 8089,
	CY Cergy Paris Universit\'e, 95302 Cergy-Pontoise Cedex, France}

\author{Stefano Scopa}
\affiliation{Laboratoire de Physique Th\'eorique et Mod\'elisation, CNRS UMR 8089,
	CY Cergy Paris Universit\'e, 95302 Cergy-Pontoise Cedex, France}
\affiliation{SISSA and INFN, via Bonomea 265, 34136 Trieste, Italy}

\author{Jacopo De Nardis}
\affiliation{Laboratoire de Physique Th\'eorique et Mod\'elisation, CNRS UMR 8089,
	CY Cergy Paris Universit\'e, 95302 Cergy-Pontoise Cedex, France}

\date{\today}

\begin{abstract}
	We consider one-dimensional interacting quantum fluids, such as the Lieb-Liniger gas. By computing the low-temperature limit of its (generalised) hydrodynamics we show how in this limit the gas is well described by a conventional viscous (Navier–Stokes) hydrodynamics for density, fluid velocity and the local temperature, and the other generalised temperatures in the case of integrable gases. The dynamic viscosity is proportional to temperature and can be expressed in a universal form only in terms of the emergent Luttinger Liquid parameter $K$ and its density. We show that the heating factor is finite even in the zero temperature limit, which implies that viscous contribution remains relevant also at zero temperatures.   Moreover, we find that in the semi-classical limit of small couplings, kinematic viscosity diverges, reconciling with previous observations of Kardar-Parisi-Zhang fluctuations in mean-field quantum fluids. 
\end{abstract}

\maketitle

\prlsection{Introduction}
Quantum many-body interacting systems pose an immense technical challenge to modern-day physics due to their exponential complexity. A successful approach in the past years has been in borrowing concepts and ideas from classical hydrodynamic theory and applying them to quantum systems \cite{Peng2023,PhysRevFluids.5.104802,1805.09331,Lucas2018,PhysRevB.105.205127,PhysRevLett.103.216602}. The main idea behind these methods is the same in classical and quantum physics: the exponentially large information determining the state of the system is reduced to few thermodynamic functions, the hydrodynamic fields, that well characterize the local equilibrium state. For quantum gases in one spacial dimension, 
the theory of generalised hydrodynamics (GHD) \cite{castro2016emergent,bertini2016transport} (see also e.g.~\cite{doyon2020lecture,alba2021generalized,essler2022short} for reviews) has shown to be able to perfectly capture the dynamics of integrable e.g.~\cite{doyon2017large,doyon2018soliton,caux2019hydrodynamics,bertini2019transport,doyon2017note,bastianello2018generalized,bertini2018entanglement,alba2019entanglement} and near-integrable quantum gases \cite{bastianello2020generalized,bastianello2021hydrodynamics,2212.12349,PhysRevX.12.041032,PhysRevLett.126.090602,Malvania_2021}, as well as spin chains e.g.~\cite{10.21468/SciPostPhys.6.1.005,piroli2017transport,bulchandani2017solvable,bulchandani2018bethe,collura2018analytic,dupont2020universal,gruber2019magnetization,scopa2022exact}, Fermionic systems \cite{mestyan2019spin,scopa2021real,scopa2022generalized,nozawa2020generalized,nozawa2021generalized,PhysRevLett.128.190401} and classical field theories \cite{2305.10495,Koch2022,Bonnemain2022,Doyon2017}. At small values of temperatures and for gapless or gapped systems, GHD recovers historically well  established results on the dynamics of low-temperature systems, in particular the celebrated Luttinger liquid theory \cite{haldane1981effective,haldane1981luttinger,giamarchi2003quantum,cazalilla2004bosonizing,PhysRevLett.102.126405,Imambekov2009,RevModPhys.84.1253} and the semi-classical approaches \cite{PhysRevLett.78.943,PhysRevB.57.8307}.  
In particular,  by re-quantising the fluctuations on top of a classical background, evolving with GHD at zero temperature,  one recovers a Luttinger Liquid theory on top of an evolving hydrodynamic fluid, connecting this way GHD with the most relevant field theory description of one dimensional quantum systems \cite{ruggiero2020quantum,collura2020domain,scopa2021exact,scopa2022exact,ruggiero2021quantum,scopa2023one}. 

At zero temperature (or entropy) GHD is effectively a set of equations describing the dynamics of the Fermi points of the fermionised degrees of freedom \cite{doyon2017large}. In particular, it was shown that, at the level of Euler hydrodynamics, therefore neglecting any viscosity, the GHD evolution of a single Fermi sea (with two Fermi points) is exactly equivalent to the conventional hydrodynamics (CHD) evolution of density $\rho$ and momentum (or fluid velocity $\eta$) as the two relevant hydrodynamic modes \cite{doyon2017large,schemmer2019generalized}. However, while the latter creates hydrodynamic shocks and fails to be meaningful after the time $t^*$ of the creation of the first shock \cite{bettelheim2006orthogonality,bettelheim2012quantum,bettelheim2006nonlinear,simmons2020quantum}, Euler GHD is free of shocks as at time $t^*$ two new modes, i.e. two new Fermi points are created, resolving this way the shock into a simple contact singularity \cite{doyon2017large}. However, since Euler GHD is valid for strictly integrable systems, it is unclear what is instead the correct hydrodynamics for quasi or even non-integrable gases at very low temperatures, that can describe different experimental settings. 

In this letter, we show that as Euler CHD is plagued by hydrodynamic shocks, one, on the other hand, cannot neglect viscosity terms that are finite as soon as interparticle interactions are non-zero. Viscous, or diffusive terms, have been indeed incorporated into GHD for a few years \cite{de2018hydrodynamic,de2019diffusion,durnin2021diffusive,medenjak2020diffusion,de2023hydrodynamic}. They have been shown to be essential for the thermalisation of quasi-integrable systems \cite{bastianello2020thermalization} and for well-describing spin dynamics in integrable spin chains \cite{de2019anomalous,de2020universality,de2021stability,de2020superdiffusion,de2022subdiffusive,bulchandani2021superdiffusion}. However, since integrable systems are typically ballistic, they usually account for small perturbative effects on top of the ballistic current. Here we shall show that the picture changes drastically at very low temperature: when $T\to0$ these terms enter the CHD as a \textit{dynamic viscosity} 
$\mu(\rho,T)$ and fully regularize its shocks, making the resulting viscous CHD a perfectly valid hydrodynamics for low-temperature gapless systems. By taking a low-temperature limit of GHD we here determine a simple and universal expression for the dynamic viscosity 
which only depends on the density $\rho$ and on the Luttinger liquid parameter $K(\rho)$ for a given interaction strength. Therefore, we claim that our result is universal for any one-dimensional interacting system at low temperatures.

\prlsection{GHD and CHD}
We start by deriving CHD by taking the low-temperature limit of GHD, and we consider the Lieb-Liniger model \cite{lieb1963exact} as a reference, although our derivation is fully generic. The Lieb-Liniger model for $N$ contact-interacting bosonic particles in an external potential $V(x)$ is given by the Hamiltonian
\begin{equation}\label{eq:lieb-liniger-H}
{
    \hat{H} = - \sum_{i=1}^N \frac{\hbar^2}{2m}\partial_{x_i}^2 +  \frac{\hbar^2c}{m} \sum_{i<j=1}^N \delta(x_i-x_j) + \sum_{i=1}^N V(x_i)}
\end{equation}
and it represents a paradigmatic model for one-dimensional interacting systems and cold atomic gases, see e.g. Ref.~\cite{cazalilla2011one,PhysRevLett.100.206805,PhysRevLett.130.020401}. { For convenience, we consider from thereafter $\hbar=m=1$, and $k_B=1$ to express temperature in units of $m k_B/\hbar^2$.} In the repulsive regime $c>0$, its eigenstates are labelled by (fermionic) quasiparticles with bare energy $\varepsilon(\theta) = \theta^2/2$ and momentum $k(\theta)= \theta$, where $\theta \in [-\infty,\infty]$ are the rapidities or the quasi-momenta of the particles.  In the thermodynamic limit, the state is specified by a filling function $n(\theta)$ fixed by the temperature and chemical potential. Within the framework of GHD, one assumes that at each position $x,t$ there exists a fluid cell where the gas is locally thermodynamic, and where we can introduce a local filling function $n(\theta;x ,t )$. The time evolution of the latter then reads, given also the external force $\mathfrak{f} = - \partial_x V$, as 
\begin{equation}\label{eq:GHD}
    \partial_t n(\theta; x ,t) + v^{\rm eff}(\theta; x ,t ) \partial_x  n(\theta; x ,t) + \mathfrak{f} \partial_\theta n(\theta; x ,t) = 0
\end{equation}
where $v^{\rm eff}(\theta; x ,t ) =  (\partial_\theta\varepsilon)^{\rm dr}/(\partial_\theta k)^{\rm dr}$ is the dressed velocity of the quasiparticles on top of the background fixed by the filling $n(\theta; x ,t)$. The latter can be found by knowing the explicit dressing operation, which reads as $f^{\rm dr} = (1- \varphi n )^{-1} \cdot f $ with the scattering matrix $\varphi(\theta) = c/(\pi (c^2 + \theta^2))$ acting as a convolution operator $\varphi  \cdot  f  = \int \varphi (\theta - \alpha) f(\alpha) d\alpha$.  

At low temperatures, the filling function becomes very close to a sharp Fermi sea, such that $\de_x n(\theta;x,t)=- \sum_{\sigma} \sigma \de_x\theta^\sigma(x,t) \delta(\theta - \theta(x,t)^\sigma)$ with $\theta^\sigma$ the two Fermi edges, indexed by $\sigma = \pm 1$, and  Eq. \eqref{eq:GHD} can be rewritten as an equation for the two Fermi edges~\cite{doyon2017large}
\begin{equation}\label{eq:FermiEdges}
  \sum_\sigma \delta(\theta - \theta^\sigma) [ \partial_t \theta^\sigma + (\eta + \sigma v_F(\rho) ) \partial_x \theta^\sigma -\mathfrak{f}]=0.
\end{equation}
For convenience, we described the fluid in a co-moving frame with fluid velocity $\eta = (\theta^+ + \theta^-)/2$, such that at each position $x$ the gas is found in a Fermi sea with symmetric Fermi edges $\pm q=\pm (\theta^+ -\theta^-)/2$, and Fermi velocity $v_F = \pi \rho/K$ proportional to the local density $\rho$,  with $K$ the Luttinger liquid parameter,  { obtained from the dressed charge as $K\equiv 1^{\rm dr}(\theta^\pm)^2$ \cite{korepin1997quantum,giamarchi2003quantum}}. Eq.~\eqref{eq:FermiEdges} can be shown, see Supplementary Material (SM) \cite{SM}, to be fully equivalent to CHD for density and fluid velocity readings as
\begin{eqnarray}\label{eq:zeroCHD}
    \partial_t \rho  &+& \partial_x (\eta \rho) =0\nn\\
    \partial_t\eta &+& \eta \partial_x \eta + \rho^{-1} {\partial_x   \mathcal{P}_s^{(0)}(\rho) }{ } =  \mathfrak{f}, 
\end{eqnarray}
with $ \mathcal{P}_s^{(0)}(\rho)$ is the static pressure of the gas at zero temperature, fixed by the equation of state of the system. In the specific example of the Lieb-Liniger gas,  the static pressure at zero temperature is given by the integral of the dressed energy $\varepsilon^{\rm dr}$, $ \mathcal{P}^{(0)}_s(\rho) = - \int_{-q}^{q} d\theta \varepsilon^{\rm dr}(\theta)/2\pi $, and where $q$ is obtained by inverting the relation $\rho = \int_{-q}^{q} 1^{\rm dr}(\theta) d\theta/2\pi$. Notice also that the fluid velocity $\eta$ is related to the momentum field $p$ as $\eta=p/\rho$.
 
The equation for the fluid velocity $\eta$ in eq.~\eqref{eq:zeroCHD} takes the form of the celebrated Burgers equation, which is known to display hydrodynamic shocks \cite{PhysRevLett.125.180401}. In Euler GHD such shocks are resolved by adding to eqs.~\eqref{eq:FermiEdges} two extra Fermi edges, i.e. by introducing $2n$ hydrodynamic fields $\theta^{\sigma}$ with $\sigma=1,\ldots,2n$ that describe $n$ split Fermi seas \cite{doyon2017large}. In the phase-space $(\theta,x)$ this translates into the existence of a  \textit{continuous} one-dimensional contour $\Gamma(s)$ that separates the region where $n(\theta,x)=1$ from the region where $n(\theta,x)=0$.

Clearly, this picture is based on the underlying integrability of the model, i.e. in the fact that all the rapidities $\theta$ are conserved quantities. For non-integrable generic systems, it is clear that this phenomenon cannot be the one that regularises the shocks. As shocks in the Burgers equation are regularised by a finite viscosity, we shall then see in the coming section how to indeed introduce viscosity in CHD \eqref{eq:zeroCHD}.

\prlsection{Viscous CHD}
We now wish to include diffusive or viscosity effects. In order to do so, we shall consider local thermodynamic states with finite but small temperature, as there are no viscosity effects at strictly zero temperature. We consider therefore local canonical equilibrium states with given density $\rho$ (or chemical potential $\mu$), fluid velocity $\eta$ and temperature {$ T= 1/\beta\ll T_d$, with $T_d=\rho^2/2$ the quantum degenerate temperature of the gas \cite{Bouchoule2011}}. In the Lieb-Liniger model, this corresponds to filling functions of the form $n(\theta; x , t ) = (1 + e^{\beta \varepsilon^{\rm dr}(\theta; x ,t )})^{-1}$ given in terms of its pseudo-energy
\begin{equation}\label{eq:TBA}
     \varepsilon^{\rm dr} = \varepsilon(\theta - \eta(x,t)) - (q(x,t))^2 - \frac{\varphi  \cdot \log (1+ e^{- \beta  \varepsilon }) }{\beta(x,t)  }.
\end{equation}
We then move to introduce diffusion terms into CHD by taking the high $\beta$ limit of the diffusive GHD, which is known from refs.~\cite{de2018hydrodynamic,de2019diffusion} and it is given by adding to the r.h.s. of eq. \eqref{eq:GHD} the diffusive part 
\begin{equation}
    R  \cdot \partial_x (R^{-1} \cdot\tilde{D} \cdot  \partial_x n (\alpha)) ,
\end{equation}
with the kernels $R_{\theta,\theta'}$ and $\tilde{D}_{\theta,\theta'}$ presented in \cite{SM}.  
As diffusive GHD contains $\partial_x^2$ terms, a description only in terms of the Fermi edges becomes impossible, as we cannot simply use that the derivative of the filling is a delta function at the edges, as this would produce $\delta'(\theta  -  \theta^\sigma) = \partial \delta (\theta - \theta^\sigma) $ whose meaning outside integration is unclear.

The only way to clarify their role is to plug them within the hydrodynamic equation for the only two modes which are relevant here, i.e. density and momentum (fluid velocity). This already clarifies the essential difference of viscous CHD from Euler GHD: while the latter can be simply obtained as an equation for the evolution of the Fermi edges \eqref{eq:FermiEdges} it can be trivially extended to any number of them during time evolution. Viscous CHD instead requires fixing the number of Fermi edges from the start, and, as we shall see, does not require producing new Fermi edges, as hydrodynamic shocks are regularised by viscosity. By means of a long but straightforward calculation deferred to the SM \cite{SM}, we then arrive to 
\begin{eqnarray}\label{eq:viscousGHD}
    \partial_t \rho  &+& \partial_x (\eta \rho) =0,\nn\\
    \partial_t\eta &+& \eta \partial_x \eta  + \frac{\partial_x  \mathcal{P}_s(\rho,T) }{\rho }= \frac{\partial_x(  \mu(\rho,T) \partial_x \eta )}{\rho} + \mathfrak{f},
\end{eqnarray}
with the dynamic viscosity at the leading order is linear in temperature, as already remarked previously \cite{PhysRevA.107.013310}. In terms of {dimensionless interaction $\gamma=c/\rho$ and temperature $\tau=T/c^2$, the dynamic viscosity reads as  
\begin{equation}
\label{eq:viscosity}
   \frac{\mu(\gamma,\tau)}{c}  =    \frac{K \tau \gamma^3}{4\pi} (\de_\gamma \log K)^2 + {\cal O}(\tau^3),
\end{equation}
which can be also rewritten for fixed interaction $c$ as $ \mu(\rho,T)  =\frac{\rho T K}{4\pi}(\de_\rho \log K)^2 =T{ \left(1-K \partial_\rho v_F /\pi \right)^2} /{4 v_F}$.} This result can be understood similarly as in the kinetic picture of diffusion in integrable models proposed in \cite{PhysRevB.98.220303}: the excitations close to the Fermi edges move with Fermi velocity $v_F(\rho) = \pi \rho/K(\rho)$ but as there are local density fluctuations $\rho \to \rho + \delta \rho$ within the fluid, the Luttinger parameter (i.e. the Fermi velocity) also fluctuates $K \to K + \partial_\rho K \delta \rho $, giving a diffusive spreading to the trajectories. The dependence of the Luttinger parameter on density (and therefore on the position in an inhomogeneous fluid) is what makes inhomogeneous Luttinger liquids for interacting systems non-conformal invariant \cite{Bastianello2020} and it is a direct effect of non-trivial interactions. {In the so-called Tonks regime of strong repulsion $\gamma\gg 1$, the dependence of the Luttinger parameter on the density trivialises, $K \to 1$, and viscosity vanishes, as expected for non-interacting particles. The first correction at $\gamma\gg 1$ is obtained from \eqref{eq:viscosity}  using $K\simeq (1+4/\gamma)$ \cite{,cazalilla2004bosonizing}, and reads as
 \begin{equation}
     \frac{\mu(\gamma,\tau)}{c} \simeq  \frac{4  \tau }{\pi\gamma} + {\cal O}(\gamma^{-3}).
 \end{equation}
 }
{The pressure in \eqref{eq:viscousGHD} of the interacting gas} is given by taking the first correction to static pressure in $T^2$, $\mathcal{P}^{(0)}_s(\rho) \to \mathcal{P}_s(\rho,T)$ 
with 
\begin{equation}
     \mathcal{P}_s(\rho,T) =  \mathcal{P}^{(0)}_s(\rho) + T^2 \rho \tilde{\chi}_e /2 ,
\end{equation}
with $ \tilde{\chi}_e=  {K}/({3  \rho^2})$. 
Eq. \eqref{eq:viscousGHD} gives the evolution of density and momentum of the fluid, which is one-to-one with the chemical potential $q$ and the boost $\eta$. Given that the system is at a finite temperature, and this also represents a hydrodynamic variable, an extra equation is needed: the one for the evolution of the energy density $e$. We define energy density at rest as $e = E/\rho - \eta^2/2$, 
where $E$ is the total energy of the fluid, which in the Lieb-Liniger model is computed via $E = \int d\theta   n(\theta) (\theta^2/2 + V(x))^{\rm dr}$. By proceeding in an analogous manner as to derive the eqs. \eqref{eq:viscousGHD}, we obtain 
\begin{equation}
    \partial_t e + \eta \partial_x e + \frac{\mathcal{P}_s(\rho,T)}{\rho} \partial_x \eta  =  \frac{ \mu(\rho,T)}{\rho} (\partial_x \eta)^2    . 
\end{equation}
As expected, the \textit{kinematic viscosity} $\nu = \mu/\rho$ now enters the equation. We can convert this equation into an equation for the evolution of the temperature field $T(x,t) = 1/\beta(x,t)$ using the definition of the energy susceptibility at fixed density  $\delta e/\delta T \Big|_{\rho}  = \tilde{\chi}_e T$. Given the expression at zero temperature for generic interacting systems, we then obtain  
\begin{eqnarray}\label{eq:GHDEnergy}
  \partial_t T &+& \eta \partial_x T  = -  \frac{\mathcal P^T_s}{\rho \tilde \chi_e} T \partial_x \eta  +  \frac{\mu(\rho,T)}{T \rho  \tilde{\chi}_e } (\partial_x \eta)^2.
\end{eqnarray}
On the r.h.s. of \eqref{eq:GHDEnergy}, $ \mathcal P^T_s  =  \pi \rho \left(1/\rho + {\partial_\rho v_F} /v_F \right)/(3v_F ) $ is the low temperature correction to the stationary pressure at constant density, and we omitted the thermal conduction $\partial_x( \kappa(\rho,T)   \partial_x T) $ with $ \kappa(\rho,T) \sim T^2 K \mu(\rho,T) [\partial_\rho (\tilde \chi_e K)]^2/32$, which is subleading in temperature. Both terms are derived in the SM \cite{SM}. 

Eqs. \eqref{eq:viscousGHD} and \eqref{eq:GHDEnergy}  take exactly the same form as the standard Navier-Stokes equations for a fluid, i.e. continuity equation, conservation of momentum and conservation of energy, and it is quite remarkable that we can derive them in an exact, non-perturbative way.
\begin{figure}[!t]
    \centering
      \includegraphics[width=\columnwidth]{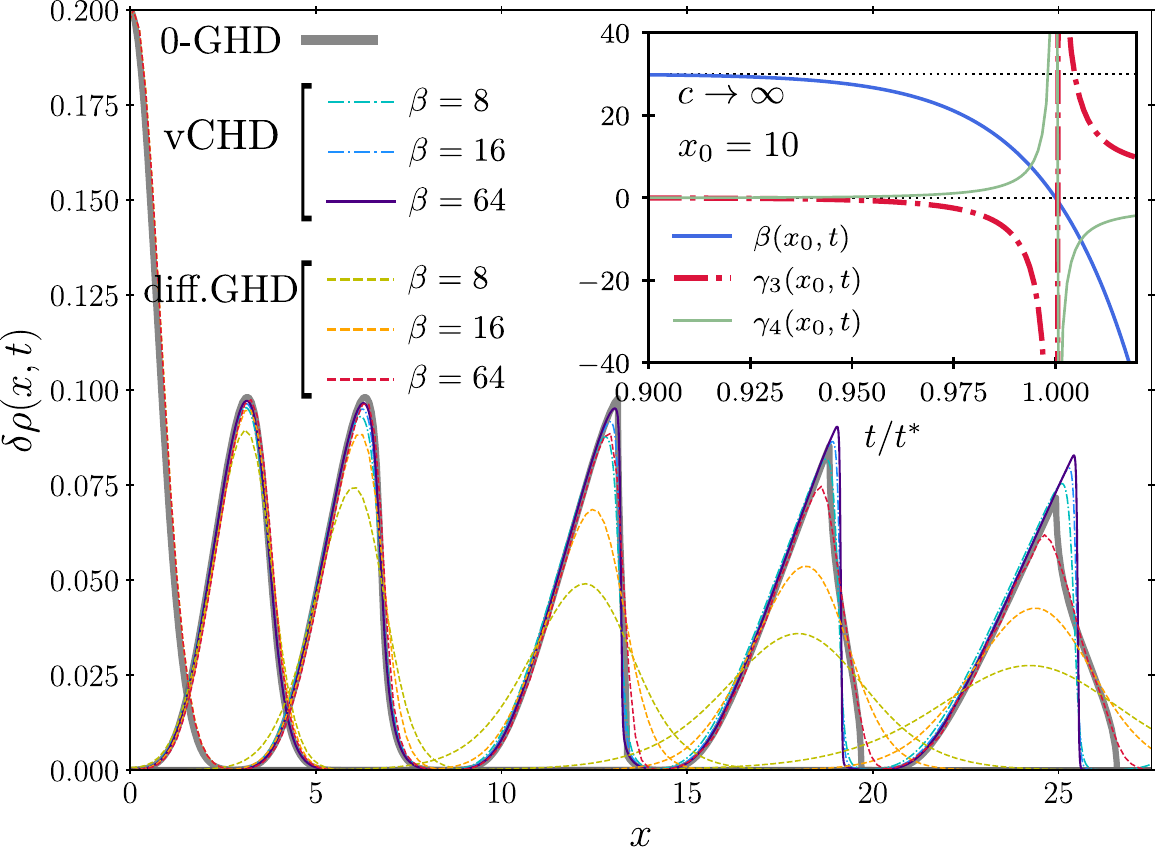}
    \caption{Evolution of the excess density $\delta\rho=\rho-\rho_{\infty}$ for an initial bump obtained as the ground state density of \eqref{eq:lieb-liniger-H} with interaction $c=1$ and Gaussian potential $V(x)=-a_1-a_2 \exp(-x^2/\sigma^2)$,  released at $t>0$; here $\sigma=1$, $a_{1,2}$ are set such that the background density $\rho_{\infty}=1$ and maximum $\rho_0=1.2$. We compare the result by zero-entropy GHD (0-GHD, solid thick lines), diffusive GHD (dashed lines) and viscous CHD (vCHD, dash-dotted lines) at different values of inverse temperatures $\beta$ (see legend).{ $K(\gamma\equiv 1)\simeq 3.4$ is obtained from Bethe ansatz \cite{korepin1997quantum} and used in \eqref{eq:viscosity}}. Times are $t=0,3,6,12.5,18.5,25$ and increase from leftmost to rightmost peak in the figure. \emph{Inset}~--~Evolution of the inverse temperature and generalised temperatures starting from a thermal case at low temperature in the integrable gas, here for simplicity the Lieb-Liniger gas at large coupling (Tonks limit). The temperatures are obtained by fitting the quasiparticle occupations $n(\theta;x ,t)$ at a given position $x_0$ during the front dynamic expansion. We clearly see the moment when the Fermi sea split as the moment of temperatures inversion. 
}
    \label{fig:viscousCHD_vs_GHD}
\end{figure}
We first notice that the heating factor in eq. \eqref{eq:GHDEnergy} is $\sim \mu/T$ namely it is order zero in temperature. Therefore, even if the system is initially at zero temperature, the rapid growth of the velocity gradient $\partial_x \eta$ heats the system, giving therefore finite viscosity $\mu$ to the dynamics of $\eta$ and regularising its shocks. 
Moreover, given that the dynamic viscosity in eq.~\eqref{eq:viscosity} is expressed only in terms of the universal features of the low-temperature effective field theory, namely the Luttinger liquid constant $K$ and its density-dependence (which is non-trivial only for interacting quantum gases), we conjecture that its form is universal for generic one-dimensional quantum fluids. The argument is simple: at low temperatures, all quantum fluids become effectively integrable, as their description is in terms of Luttinger liquid freely propagating bosonic modes. Their diffusion is therefore expected to be the same as for integrable quasiparticles, where the kinetic picture explained above applies.   We also stress that there is no fundamental difference in the nature of transport at small temperatures between integrable and non-integrable systems, contrary to what is claimed in previous literature \cite{PhysRevLett.119.036801}.


 
\prlsection{Front dynamics in integrable gases}
We now focus on the fate of an initial density bump, for example in a system at a given temperature {$T_0\ll T_d$} and density $\rho_0$. Such a setting is paradigmatic to understanding the response of a system to external perturbations, and we here use it to establish the main differences between viscous CHD and integrable diffusive GHD at low temperature, see Fig.~\ref{fig:viscousCHD_vs_GHD}. {When the system is integrable, one could expect that not only temperature and chemical potential can characterise a local stationary state, but a large number of Lagrange multipliers $\gamma_n$ associated to higher conserved quantities $\langle\hat{Q}_n\rangle= \int d\theta n(\theta) (\theta^n)^{\rm dr}/n!$, i.e.~a Generalised Gibbs Ensembles (GGE), deviating this way from the behaviour of a generic non-integrable system. This amounts to replacing the bare energy $\varepsilon=\theta^2/2$ with a higher-order polynomial
\begin{equation}\label{eq:new-bare-en}
    \varepsilon= \frac{\theta^2}{2}   +\sum_{n\geq 3} \gamma_n(x,t)  \frac{\theta}{n!}
\end{equation}
in eq.~\eqref{eq:TBA}, still yielding a valid stationary state of the Lieb-Liniger gas, due to its integrability.
By extending the result for the temperature field in \eqref{eq:GHDEnergy} to the higher potentials $\gamma_n(x,t)$, we find quite lengthy partial differential equations, see \cite{SM} for their expression. The main relevant fact is that while $\partial_t T \sim O(T)$, we instead find $\partial_t \gamma_n \sim O(T^0)$. Namely, even if we prepare a thermal gas at low temperature (where $\gamma_{n}(x,0)=0$), the integrable gas will generate finite generalised temperatures in the post-shock dynamics. For instance, at the shock time $t^*$ for a certain position $x_0$, we find that $\beta(x_0,t)<0$ and $\gamma_{3,4}(x_0,t)\neq 0$, leading to the splitting of Fermi seas defined via \eqref{eq:new-bare-en} (cf.~inset of Fig.~\ref{fig:viscousCHD_vs_GHD}). Similarly, higher potentials are activated when the Fermi seas further split. Such temperature dynamics leads to a significantly different shock regularisation in GHD compared to the CHD one, even if both are regular hydrodynamics. Indeed, while GHD displays a growing shock region, viscous CHD converges to a given profile, with a finite front width, and it never develops a shock as temperature is lowered. Namely, even if temperature $T\to 0$, the heating factor $\mu/T$ entering \eqref{eq:GHDEnergy} remains finite, and leads to a self-regulation of the shock driven by the kinematic viscosity.}
 
  In Fig.~\ref{fig:viscousCHD_vs_GHD}, zero-temperature GHD agrees with diffusive GHD at finite temperature, signalling how the zero-temperature approximation is often able to capture out-of-equilibrium fluids. On the other hand, observed deviations from viscous CHD are attributed to the absence of higher conservation laws, as discussed above.

\prlsection{The small coupling limit $c\to0^+$ and KPZ physics}
 As already discussed in the first Lieb-Liniger paper \cite{lieb1963exact}, the limit of small coupling of the Lieb-Liniger gas does not simply recover free bosons. Indeed, when  $c \to 0^+$, its ground state becomes the one describing the so-called (semi-classical) condensate solution of the Nonlinear Schrödinger equation (NLS) \cite{Ishikawa1980}. This is characterised by vanishing Fermi momentum $q \sim \sqrt{c}$ but with diverging dressed functions $1^{\rm dr} \sim 1/\sqrt{c}$ in order to keep the density $\rho$ finite in the limit. The relevant excitations become then the Bogoliubov excitations with spectrum given by $\varepsilon_k \sim |k|$ and therefore with a degenerate group velocity $v_k \sim {\rm sgn}(k)$. As $\partial_\rho v_F(\rho) \sim \lim_{q \to 0} \partial_q v_q $ the latter diverge, giving a divergent dynamic viscosity at low coupling as 
\begin{equation}
    \mu (\rho,T) \sim \frac{T}{\sqrt{c}} ,
\end{equation}
signalling the breakdown of the viscous CHD and the emergence of Kardar-Parisi-Zhang physics \cite{PhysRevLett.56.889}. The latter is well-known to emerge in the stochastic Burgers equation, i.e. given a white $\delta-$correlated noise $w$, a local perturbation of a hydrodynamic field $\phi$ satisfying  
$
    \partial_t \phi + \partial_x\left( v \phi + \kappa    \phi^2 + w \right)=0,
$
moves with finite velocity $v$ and \textit{spreads super-diffusively}, as opposed to the diffusive case of eq.~\eqref{eq:viscousGHD} whenever $\mu$ is finite. Such a phenomenon is known to appear in generic finite-components one-dimensional fluids, as described by the non-linear fluctuating hydrodynamic (NLFH) theory \cite{Spohn2014}, which can be successfully applied also to lattice (i.e. non-integrable) NLS \cite{Mendl2015}. Again, the argument is simple: whenever the Euler currents contain non-linearities, one should expect that the introduction of a small noise (which for example can describe the interaction with other non-hydrodynamic modes) always leads to the KPZ universal fixed point. However, this is not the case at finite coupling since, although it is true that eq.~\eqref{eq:viscousGHD} contains the non-linearity $\eta^2$, there exists a continuous, thermally activated, spectrum of modes around the {Fermi points with velocities $v_F \pm \delta v(\theta)$ (with $\delta v(\theta) \ll v_F$ peaked at $\theta=\theta^\sigma$)}  that are responsible for a finite diffusion constant in the system.  It is only in the small coupling limit $c\to 0^+$ that the velocities of all such excitations become degenerate, therefore diminishing the number of effective hydrodynamic modes to only the three macroscopic ones. In this limit, therefore, the theory of NLFH applies and KPZ physics emerges, as signalled by the divergent dynamic viscosity. We should stress that the existence of KPZ physics in the NLS at low temperature was first established in \cite{Kulkarni2013} and we here give a first analytical proof of its divergent diffusion constant.  Moreover, it is interesting to notice that the degeneracy of the hydrodynamic modes leads to super-diffusive KPZ physics similarly also in the Heisenberg spin chain (and any other integrable model with non-abelian symmetry) \cite{PhysRevLett.122.210602,bulchandani2021superdiffusion} at finite temperatures. There the relevant degenerate excitations are not the ones around the Fermi points but the so-called giant magnons \cite{PhysRevLett.125.070601,2212.03696}, namely magnonic excitations with large spin and vanishing velocities.

\prlsection{Conclusion}
We have here shown that the standard Navier-Stokes equations for the evolution of density, momentum and temperature can be derived from the low-temperature expansion of the GHD for the Lieb-Liniger gas. We have found universal expressions for the linear part  in temperature of the dynamic viscosity, which we conjecture to apply to generic one-dimensional fluids. We have shown that the viscosity is zero in the free fermionic limit, as expected, and that it diverges in the semi-classical limit of weakly interacting bosons at small temperatures, which despite many numerical works, it was never established from first principles. The divergent viscosity signals the emergence of KPZ super-diffusive spreading \cite{Kulkarni2013}, in analogy to the one observed in integrable spin chains \cite{Wei2022}.

We have shown that the viscous terms regularise hydrodynamic instabilities in one-dimensional gases, although the inclusion of generalised temperatures is necessary in order to predict the full form of the shock front in the integrable gas. 
{Moreover, one should also expect that when the system is strongly out of equilibrium and thus gradients of $\eta$ become large at the shocks points, the system strongly heats locally, invalidating, therefore, the zero-entropy approximation.} Our findings therefore suggest that zero-entropy hydrodynamics becomes harder to physically realise whenever interactions are present.

Differently from previous attempts to derive viscosity in quantum fluids by perturbative corrections to Luttinger liquids, see for example \cite{PhysRevLett.103.216602,PhysRevLett.119.036801,PhysRevA.107.013310,PhysRevLett.125.076601}, we here derive non-perturbatively using the generalised hydrodynamic of a specific model, the Lieb-Liniger gas, and we extend our result to generic systems, given the universality of its formulation. Clearly, a different derivation only involving Luttinger liquid modes would also be desirable in the future.   Our result is ready to be checked by means of numerical simulations \cite{2303.00663,PhysRevLett.121.090603,PhysRevResearch.3.013078} and to apply to different quantum fluids as such as chiral edge modes \cite{PhysRevB.41.12838,WEN1992,PhysRevA.107.033320}, to open the way for a full fluctuating hydrodynamic theory \cite{2304.03236} of Luttinger liquid field theories, and to unveil Burgers-like turbulent phases \cite{PhysRevLett.129.114101} in low-dimensional quantum fluids. 

\prlsection{Acknowledgements}
 We are thankful to B. Doyon, K. Kheruntsyan, I. Bouchoule, and J. Dubail for insightful discussions. We are in debt to M. Panfil for sharing his notes on dressed kernel identities. We thank G. Del Vecchio Del Vecchio for the related collaboration.  
 This work has been partially funded by the ERC Starting Grant 101042293 (HEPIQ) (J.D.N., S.S and A.U.) and by the ERC Consolidator
Grant 771536 (NEMO) (S.S.). S.S. is thankful to LPTM (Cergy-Paris) for the kind hospitality during the development of this work.
\bibliography{apssamp}

\onecolumngrid
\newpage

\appendix
\setcounter{equation}{0}
\setcounter{figure}{0}
\renewcommand{\thetable}{S\arabic{table}}
\renewcommand{\theequation}{S\thesection.\arabic{equation}}
\renewcommand{\thefigure}{S\arabic{figure}}
\setcounter{secnumdepth}{2}

\begin{center}
{\Large Supplementary Material \\ 
\titleinfo
}
\end{center}
\tableofcontents

\appendix

\section*{Outline}
This supplementary material is arranged as follows: \\

In App.~\ref{app:lieb_liniger_app} the Lieb-Liniger model is used to introduce our local states by use of the thermodynamic Bethe ansatz (TBA). This section includes a review of the dressing relation, which appears throughout the various calculations. As a warm-up $T=0$ generalized hydrodynamics (GHD) is considered in App.~\ref{app:ghd_evolution}, where relations between the hydrodynamic fields and TBA quantities are used to reproduce the known equivalence between $T=0$ conventional hydrodynamics (CHD) and GHD. The diffusive GHD equations are introduced, which are then specialized to the cases of momentum and energy. At low temperatures the diffusive part is determined first for momentum, App.~\ref{app:interacting_diffusion}, followed by energy in App.~\ref{app:energy_hd_eq}. Next, the question of additional Lagrange parameters is considered in App.~\ref{app:additional_Lagrange}. This consideration requires the derivation of additional evolution equations for the corresponding conserved charge, which leads to an evolution equation for the additional Lagrange parameters. At low temperature initial states these additional parameters are found to grow as $O(T)$, unlike temperature, which was found earlier to have corrections $O(1)$. These low $T$ diffusive corrections are considered in detail for both cases, and result in a closed set of low $T$ diffusion equations referred to as the viscous GHD equations. These equations are collected in their final form in App.~\ref{app:viscous_ghd_final}, where the large $c$ limit is noted. This is followed up with a numerical comparison between the predicted low $T$ diffusion and the finite temperature diffusion kernel corrections to the momentum in App.~\ref{app:numericalCheck_visc}, where additional details about the method can be found. Small temperature corrections to the hydrodynamic fields are evaluated in detail and reported in Apps.~\ref{app:HDfieldsT0}-\ref{app:subleading_temperature}. Various identities and useful relations for the dressed functions at small temperatures are collected in App.~\ref{app:id_list}. This includes the zero temperature behaviour of several hydrodynamic fields, a proof of an identity used to derive the low $T$ diffusion, general small temperature corrections to hydrodynamic fields, and a suite of dressed function identities. An alternative calculation for large $c$ is also considered in App.~\ref{app:tg_diff_deriv}, which takes advantage of the thermal energy being explicitly known and is found to match the large $c$ limit previously obtained. Additionally, the commutation of the small temperature limit and the spatial derivative are demonstrated in the large $c$ case.

    

\section{TBA for Lieb Liniger}
\label{app:lieb_liniger_app}

\subsection{Thermal energy and hydrodynamic fields}
\label{app:interacting_energy}
In this section the Galilean invariance of the Fermi weight is explored, specifically the consequences to be considered. This identity will allow us to move between a rapidity coordinate system with symmetric, and anti-symmetric Fermi points. For convenience these two setups are referred to as being the symmetric and anti-symmetric bases, respectively. For us the low-temperature Fermi weights are given by a thermal energy $\epsilon$, which is determined through the TBA relation in the presence of a chemical potential $\mu(q) > 0$. For the case of symmetric Fermi points, $\theta = \pm q$, the thermal energy is given by
\begin{eqnarray}
\label{eq:symmetric_thermal}
    \epsilon(\theta) &=&  \frac{\theta^2}{2} -  \mu(q) - \frac 1 \beta \int d\alpha \, \varphi(\theta -\alpha) \ln | 1 + \e^{-\beta \epsilon(\alpha)}| , \nn\\
    \epsilon(\pm q ) &=& 0.
\end{eqnarray}
A more general scenario can be obtained by considering a shift in the rapidity. Thus the symmetric thermal energy can be shifted into a non-symmetric scenario by noting that $\theta \in \{q,-q\}  =\{\theta^+,\theta^-\} -\eta$ with $\eta = \frac{\theta^+ + \theta^-} 2$. Furthermore, Galilean invariance of the system guarantees that any two non-symmetric Fermi points $\theta^+$, $ \theta^-$ are given by
\begin{eqnarray}
    \epsilon(\theta) &=& \frac{(\theta -  \eta)^2}{2} - \mu(q) - \frac 1 \beta \int d\alpha \, \varphi(\theta -\alpha) \ln | 1 + \e^{-\beta \epsilon}|,\nn\\
    \epsilon(\theta^\pm) &=& 0.
\end{eqnarray}
So the symmetric thermal energy in a local chemical potential $\mu(q)$ is determined from either of the above integral equations and either are used to define the Fermi weight as
\begin{eqnarray}
    n(\theta) = \left[ 1 + \e^{\beta \epsilon(\theta)} \right]^{-1}.
\end{eqnarray}
Note that a Galilean shift of the rapidity, $\theta \to \theta - \eta$, leaves the Fermi weight invariant. At $\beta \to \infty$ with $\mu(q) > 0$ the Fermi weight becomes a step function and satisfies the identity
\begin{eqnarray}
\label{eq:zero_T_fw}
    \partial_\theta n(\theta) = - \sum_{\sigma = \pm } \sigma \delta(\theta -\theta^\sigma) + O\left(\beta^{-2}\right).
\end{eqnarray}
From the Fermi weight hydrodynamic fields are introduced as
\begin{eqnarray}
    \langle Q_j \rangle = \int d\theta \, n(\theta) \dr{\left( \frac {\theta^j}{j!}\right)},
\end{eqnarray}
where an overall factor $1/2\pi$ is reintroduced in the final step and the dressed functions $\dr f$ will be defined below.

\subsection{Dressing relation}
\label{app:galilean_transform_dress}

From the the definition of the thermal energy it is possible to identify the derivative of the thermal energy with the dressed energy derivative. For Lieb-Liniger with symmetric Fermi points this explicitly reads as
\begin{eqnarray}
    \partial_\theta \epsilon = \theta + \int d\alpha \, \varphi(\theta-\alpha) n(\alpha) \partial_\theta \epsilon(\alpha) = \theta + \int d\alpha \, \varphi(\theta -\alpha) n(\alpha) \dr{\theta}(\alpha)  =: \dr \theta(\theta).
\end{eqnarray}
More generally this type of integral relation is referred to as a dressing relation.  This relation is used in conjunction with a driving function $f(\theta)$ to define a dressed function $\dr f(\theta)$ as
\begin{eqnarray}
    \dr f(\theta) = f(\theta) + \int d\alpha \, \varphi(\theta-\alpha) n(\alpha) \dr f(\alpha).
\end{eqnarray}
With the note that a left and right dressing is exactly equivalent, which implies
\begin{eqnarray}
\int d\theta    \dr f(\theta) n(\theta) g (\theta) = \int d\theta     f(\theta) n(\theta) \dr g (\theta).
\end{eqnarray}
It is also possible to dress multivariate functions like the scattering kernel
\begin{eqnarray}
    \dr \varphi(\theta,\mu) &=& \varphi(\theta - \mu) + \int d\alpha \, \varphi(\theta-\alpha) n(\alpha) \dr \varphi (\alpha,\mu),
\end{eqnarray}
with a note that once again a left or right dressing are exactly equivalent. 

There are also useful composite variables such as the effective velocity of the quasi-particles
\begin{eqnarray}
    v^{\rm eff}(\theta) = \frac{\dr \theta(\theta)}{\dr 1(\theta)},
\end{eqnarray}
which are characterized by these dressed functions. 

Note, however, that unlike the thermal energy these dressed quantities are not required to be invariant under Galilean boost. Instead the behaviour of the dressed function under a Galilean transformation depends on the form of the driving function $f$. For example, given Fermi points $\theta^\pm$ that are then boosted so $\theta \to \theta - \frac{\theta^+ + \theta^-}{2} = \theta - \eta$, which results in symmetric Fermi points $\pm q = \pm\left( \frac{\theta^+ - \theta^-}{2}\right)$, then the dressed functions behave as
\begin{eqnarray}
    \dr f(\theta-\eta) = \dr f_s(\theta) = \dr{(f(\theta -\eta))}.
\end{eqnarray}
Consider as an explicit example $f = \theta$ then
\begin{eqnarray}
        \dr \theta(\theta-\eta) &=&  \dr{(\theta - \eta)} =\dr \theta(\theta) - \eta \dr 1(\theta) . \nn\\
\end{eqnarray}
Furthermore, the Fermi velocity can be determined by evaluating the effective velocity at the symmetric Fermi point  $v^{\rm eff}(q) = v_F$. Note that with the Galilean invariance the general effective velocity satisfies
\begin{eqnarray}
    v^{\rm eff}(\theta^\pm) = \frac{\dr \theta(\theta^\pm) }{\dr 1(\theta^\pm)} = \frac{ \dr \theta(\pm q)}{\dr 1(\pm q)} + \eta = \pm v_F + \eta.
\end{eqnarray}

\section{Zero temperature Euler GHD and conventional hydrodynamics (CHD)}
\label{app:ghd_evolution}
Generalized hydrodynamics (GHD) determines the time evolution of conserved charges $q_i$, which at the Euler scale involves evolution equations of the form
\begin{eqnarray}
0 &=& \partial_t \langle Q_i \rangle + \partial_x \langle J_i\rangle = \partial_t \int d\theta (q_i n \dr 1 ) + \partial_x \int d\theta (q_i v^{\rm eff} n \dr 1 ) .
\end{eqnarray}
With the subscript '$i$' denoting the index of the conserved charge, the first few of which are density, $q_0 =1$, momentum $q_1 = \theta$, and energy $q_2 = \theta^2/2$. What distinguishes GHD from conventional hydrodynamics (CHD) is the presence of an infinite number of conserved charges, our goal here is to demonstrate that at $T=0$ the Euler scale GHD reduces to  CHD. This is a well known result, however it should be noted that this result implies the presence of shocks at non-zero temperatures. A similar procedure for truncating the GHD equations at finite but low-temperature will then be necessary for introducing a diffusive correction to regularize these shocks. 

\subsection{Euler CHD equations}
From GHD the CHD equations can be understood as demonstrated as emerging from a truncation of independent evolution equations, which occurs in the zero temperature limit of GHD. At the Euler scale the first three modes for GHD equations read
\begin{eqnarray}
    \partial_t \rho + \partial_x p &=&\partial_t \int d\theta (n \dr 1) + \partial_x \int d\theta \, v^{\rm eff} (n \dr 1) =0 ,\nn\\
    \partial_t p + \partial_x \mathcal{ P} &=& \partial_t \int d\theta \theta (n \dr 1) + \partial_x \int d\theta \, \theta v^{\rm eff} (n \dr 1) =0 ,\nn\\
    \label{eq:hydro_eq_euler_app}
    \partial_t E + \partial_x J_E &=& \partial_t \int d\theta \theta^2 (n \dr 1) + \partial_x \int d\theta \, \theta^2 v^{\rm eff} (n \dr 1) =0 .
\end{eqnarray}
It should be noted that when the, generally, infinite number of evolution equations in GHD truncate to a finite number the result is some form of CHD.
\\

A more conventional form of CHD can be written by noting $p = \eta \rho$ with $\eta = (\theta^+ + \theta^-)/2 $ being the fluid velocity so that the momentum equation can be recast as
\begin{eqnarray}    \partial_t p + \partial_x \mathcal P = \rho \partial_t \eta + \rho \eta \partial_x \eta + \partial_x \mathcal P_s =0,
\end{eqnarray}
where $\mathcal P_s = \mathcal P - \eta^2 \rho$ is the stationary pressure. The energy equation is likewise written instead as an evolution equation in terms of a stationary energy density 
\begin{equation}
    e = \frac E \rho - \frac{\eta^2}{2}.
\end{equation}
By use of their definition from TBA, the hydrodynamic fields for energy and energy current satisfy, where $i = t, x$
\begin{eqnarray}
        \partial_i E       &=&\eta \rho \partial_i \eta + \frac{\partial_i q}{v_F} (\mathcal P+E-\eta^2 \rho), \nn\\
    \label{eq:zero_T_hydro}
    \partial_i J_E     &=& (\mathcal P+E) \partial_i \eta + \eta v_F^2  \partial_i \rho - \eta^2 \rho \partial_i \eta + \eta \partial_i E ,
\end{eqnarray}
with the Fermi velocity $v_F=v^{\rm eff}(q)$. So the energy equation may be rewritten at low temperatures as
\begin{eqnarray}
\frac 1 \rho \left(\partial_t E + \partial_x J_E\right) &=& \partial_t e +  \frac {\mathcal P_s} \rho    \partial_x \eta + \eta \partial_x e + \frac{ e}{\rho} \left( \partial_t \rho +  \eta \partial_x \rho + \rho \partial_x \eta \right) + \eta(\partial_t \eta + \eta \partial_x \eta +v_F^2 \partial_x \rho )  .
\nn\\
\end{eqnarray}
These considerations then allow us to identify the first three GHD modes with the CHD equations:
\begin{eqnarray}
    \partial_t \rho + \partial_x(\eta  \rho) &=&0,\nn\\
    \partial_t \eta + \eta \partial_x \eta + \frac{\partial_x \mathcal P_s}{\rho} &\overset{\rm{Euler}}{=}&0  ,\nn\\
    \partial_t e +  \frac {\mathcal P_s} \rho    \partial_x \eta + \eta \partial_x e &\overset{\rm{Euler}}{=}&0.
\end{eqnarray}
To prove the equivalence of GHD and CHD it must be demonstrated that GHD truncates so that these relations, or a subset thereof, are sufficient for determining the dynamics.

 \subsection{Equivalence of Euler GHD with CHD at $T=0$}
At $T=0$ explicit identities for the hydrodynamic fields are determined from TBA. Again considering the first three modes the identities for the momentum and density equations are
\begin{eqnarray}
    \partial_i \rho &=& 2( \dr 1(\theta^+))^2(\partial_i q),\nn\\
\partial_i p 
&=& 2 (\dr 1(\theta^+))^2 \left( v_F\partial_i \eta + \eta (\partial_i q)\right),\nn \\
   \partial_i \mathcal P 
   &=& 2 (\dr 1(q))^2 \left[2 (\partial_i \eta)  \eta v_F +  \left( \eta^2 + v_F^2\right) \partial_i q\right].
\end{eqnarray}
Both of these equations imply conditions on the Fermi points $\theta^+,\theta^-$ and are more conveniently characterized by a symmetric form $\theta^\pm = \eta \pm q$ in terms of the symmetric Fermi points $q$ and the fluid velocity $\eta$. Specifically, they imply the Euler scale zero entropy GHD equations for density and momentum
\begin{eqnarray}
    0  &=&  \partial_t q + v_F \partial_x \eta + \eta \partial_x q  ,\nn\\
    \label{eq:zero_ghd_app}
    0 &\overset{\rm{Euler}}{=}& \partial_t \eta + \eta \partial_x \eta + v_F\partial_x q .
\end{eqnarray}
recalling that the Fermi velocity $v_F$, is the effective velocity evaluated at the the Fermi point. As a consequence of this result the Euler scale evolution of the symmetric Fermi point $q = \frac{\theta^+ - \theta^-}{2}$ and the fluid velocity $\eta = \frac{\theta^+ + \theta^-}{2}$ dictates the dynamics of the model. Thus, at $T=0$ GHD truncates to a finite number of equations establishing the CHD/ GHD equivalence. The precise meaning of this truncation is that the additional equations become redundant. Consider the energy evolution equation at the Euler scale, which from the identities
\begin{eqnarray}
    \partial_i E      
    &=& (\partial_i \eta) 2 \eta \dr \theta_s(q) \dr 1 + (\partial_i q)  \left[ \dr{(\theta^2_s)}(q) + \eta^2 \dr 1(q)\right] \dr 1, \nn \\
    \label{eq:zero_T_hydro_deriv}
    \partial_i J_E 
    &=&\partial_i \eta \left([\dr{(\theta^2_s)}(q) + \eta^2 \dr 1(q)]  \dr \theta(q)  + 2 \eta^2 \dr \theta_s(q) \dr 1(q) \right) + \partial_i q \left( \eta \dr 1(q) [\dr{(\theta^2_s)}(q) + \eta^2 \dr 1(q)]  + 2 \eta( \dr \theta_s(q))^2\right) .\nn\\
\end{eqnarray}
imply that the energy evolution equation at $T=0$ is equivalent to an 'additional' evolution equation for $\eta,q$. However, after some massaging find
\begin{eqnarray}
     \partial_t E + \partial_x J_E &\overset{\rm{Euler}}{=}& 0 ,\nn\\
\label{eq:energy_evn_euler}
 4 \eta \dr \theta \dr 1 \left[ \partial_t \eta + \eta \partial_x \eta + v_F \partial_x q \right] + 2 \left[ \dr{(\theta^2_s)} + \eta^2 \dr 1 \right] \dr 1 \left[ \partial_t q + v_F \partial_x \eta + \eta \partial_x q \right]  &\overset{\rm{Euler}}{=}& 0.
\end{eqnarray}
Comparing this with the momentum and density equations at $T=0$, see eq.~\eqref{eq:zero_ghd_app}, makes clear that the first term is zero, due to the momentum equation, and the second also zero, due to the density equation. Consequently, at Euler scale at $T=0$ the energy evolution equation is merely a restatement of the density and momentum evolution equations. Whenever the temperature is not strictly zero, the zero entropy GHD equations exhibit shocks, which in the literature have been resolved by considering the post shock state as having multiple Fermi seas. Mathematically, these shocks are not surprising.

\section{Low-temperature viscous momentum equation from GHD}
\label{app:interacting_diffusion}
GHD is capable of describing more than just the Euler scale dynamics, having been extended to account for diffusive and more recently dispersive dynamics. Here, the low T diffusive behaviour of GHD is considered, understanding the diffusion as a low-temperature correction to the conserved charges such as the density and momentum. The goal being to determine low-temperature viscous GHD equations, which are found at leading order to characterize a type of CHD with appropriate viscosity $\mu(\rho,\beta)$. From GHD the evolution equation of a given conserved charge, whose local density is given by $q_i$, in the presence of diffusion is given by
\begin{eqnarray}
\label{eq:app_ghd_diffusion}
\partial_t ( q_i n \dr 1) + \partial_x ( q_i v^{\rm eff} n \dr 1 ) = \frac {q_i(\theta)} 2 \partial_x\left( \int d\alpha D(\theta,\alpha) \partial_x (n \dr 1)(\alpha)\right).  
\end{eqnarray}
With the general form of the diffusion kernel $D(\theta,\alpha)$ being known from GHD as
\begin{eqnarray}
D(\theta,\alpha) &=&\int d\mu d\nu \, \theta (1 - n \varphi  )^{-1}(\theta,\mu) \dr 1(\mu) \tilde D(\mu,\nu) \frac{1}{\dr 1(\nu)} (1- n \varphi)(\nu,\alpha), \nn\\
 \tilde D(\theta,\mu) &=& - 2\pi \delta(\theta-\mu) \int d\alpha \left[ n(\alpha) (1 - n(\alpha)) (\dr 1(\alpha)) |v^{\rm eff}(\theta)-v^{\rm eff}(\alpha)| \left(\frac{\dr \varphi(\theta,\alpha)}{\dr 1(\theta)}\right)^2 \right]\nn\\
 \label{eq:app_diff_kernel}
     &&+ 2\pi n(\alpha)(1 -n (\alpha)) \left(\frac{ \dr \varphi(\theta,\mu)}{\dr 1(\theta)}\right)^2 \dr 1(\theta)|v^{\rm eff}(\theta - v^{\rm eff}(\mu)|.
\end{eqnarray}
The factor $2\pi$ appearing due to writing everything in terms of $\dr 1$. So rewritten in terms of the hydrodynamic fields the GHD equations for the density and momentum are
\begin{eqnarray}
    \partial_t \rho + \partial_x (\eta \rho) &=& 0,\nn\\
    \partial_t\eta + \frac{1}{2} \partial_x ( \eta^2) +\frac{\partial_x  \mathcal{P}_s(q,T) }{\rho} &=& \frac 1 \rho D(p)  .
\end{eqnarray}
With $D(p) =\partial_x [\theta (1 - n \varphi  )^{-1} \dr 1 \tilde D \frac{1}{\dr 1} (1- n \varphi)]$ being the diffusive contribution to the momentum from GHD. The low-temperature diffusive corrections to the conserved charges can now be determined from the full diffusion expression from GHD by expanding the $\tilde D$ operator.
 After this expansion, a more convenient form for the full diffusion kernel is found by using the identity
\begin{eqnarray}
    \int d\alpha \, (1- n \varphi)(\theta,\alpha) \partial_x(n \dr 1)(\alpha) = \dr 1(\theta) \partial_x n(\theta).
\end{eqnarray}
By specializing to a thermal state, like those described in App.~\ref{app:lieb_liniger_app}, the diffusion kernel can be written as
\begin{eqnarray}
D(\theta,\alpha) &=&\int d\mu d\nu \,  (1 - n \varphi  )^{-1}(\theta,\mu) \dr 1(\mu) \tilde D(\mu,\nu) \frac{1}{\dr 1(\nu)} (1- n \varphi)(\nu,\alpha), \nn\\
 \tilde D(\theta,\mu) &=& - 2\pi \delta(\theta-\mu) \int d\alpha \left[ \frac{\partial_\alpha n}{\beta  \partial_\alpha  \varepsilon(\alpha)} \dr 1(\alpha) |v^{\rm eff}(\theta)-v^{\rm eff}(\alpha)| \left(\frac{\dr \varphi(\theta,\alpha)}{\dr 1(\theta)}\right)^2 \right]\nn\\
     &&+ 2\pi \frac{\partial_\theta n}{\beta\partial_\theta \varepsilon(\theta)} \left(\frac{ \dr \varphi(\theta,\mu)}{\dr 1(\theta)}\right)^2 \dr 1(\theta)|v^{\rm eff}(\theta) - v^{\rm eff}(\mu)|.
\end{eqnarray}
Along with the $\partial_x n = (\partial_\theta n) \frac{\partial_x \epsilon}{ \partial_\theta \epsilon}$, resulting in
\begin{eqnarray}
\int d\alpha    \tilde D(\theta,\alpha) \partial_x n(\alpha) = \frac {2\pi} {\beta^2} \int d\alpha \left(\frac{  \dr \varphi(\theta, \alpha) }{\dr 1(\theta)}\right)^2 |v^{\rm eff}(\theta) - v^{\rm eff}(\alpha)| \left[   \dr 1(\theta) \partial_x (\beta \epsilon(\alpha)) - \dr 1(\alpha) \partial_x (\beta \epsilon(\theta))\right] \frac{\partial_\theta n}{\partial_\theta \epsilon} \frac{\partial_\alpha n }{\partial_\alpha \epsilon}.\nn\\
\end{eqnarray}
The form of the square bracketed term is simplified by noting that the spatial derivatives of the thermal energy satisfy
\begin{eqnarray}
\frac{\partial_x (\beta \epsilon (\mu))}{\partial_\mu \epsilon} v^{\rm eff}(\mu)-    \frac{\partial_x( \beta \epsilon (\theta))}{\partial_\theta \epsilon} v^{\rm eff}(\theta) =\beta \left( v^{\rm eff}(\theta) - v^{\rm eff}(\mu) \right) \partial_x \eta + \left( \frac{\epsilon(\mu)}{\dr 1(\mu)} - \frac{\epsilon(\theta)}{\dr 1(\theta)} \right)\partial_x \beta , 
\end{eqnarray}
which follows by of the spatial derivative identity~\eqref{eq:ratio_id}. Along with the additional simplification  $\frac{\partial_x \epsilon(\theta)}{\dr 1(\theta)} = \frac{\partial_x \epsilon (\theta)}{\partial_\theta \epsilon} v^{\rm eff}(\theta)$. As a consequence the momentum diffusion kernel can be written in a symmetric form as
\begin{align}
&\int  d\theta d\mu \,    v^{\rm eff}(\theta) (\dr 1(\theta))^2 \tilde D(\theta,\mu) \partial_x n(\mu) 
= \frac { 2\pi} {2\beta} \int d\theta  d\mu \,  \frac{\partial_\theta n}{\partial_\theta \epsilon} \frac{\partial_\mu n}{\partial_\mu \epsilon}  (\dr 1(\theta))(\dr \varphi(\theta,\mu))^2 (\dr 1(\mu))  \frac{\left( v^{\rm eff}(\theta) - v^{\rm eff}(\mu) \right)^4}{|v^{\rm eff}(\theta) - v^{\rm eff}(\mu)| } \partial_x \eta, \nn\\
&+ \frac{2\pi}{2\beta^2} \int d\theta d\mu \, \frac{\partial_\theta n}{\partial_\theta \epsilon} \frac{\partial_\mu n}{\partial_\mu \epsilon} (\dr 1(\theta) \dr 1(\mu)) (\dr \varphi(\theta,\mu))^2 \frac{(v^{\rm eff}(\theta) - v^{\rm eff}(\mu))^3}{|v^{\rm eff}(\theta)-v^{\rm eff}(\mu)|} \left( \frac{\epsilon(\mu)}{\dr 1(\mu)} - \frac{\epsilon(\theta)}{\dr 1(\theta)} \right) \partial_x \beta,\nn\\
=:&  \int \frac { d\theta  d\mu} {2\beta} \,  \frac{\partial_\theta n}{\partial_\theta \epsilon} \frac{\partial_\mu n}{\partial_\mu \epsilon} \left[H(\theta,\mu) \partial_x \eta + h(\theta,\mu)\left(\frac{\epsilon(\mu)}{\dr 1(\mu)} - \frac{\epsilon(\theta)}{\dr 1(\theta)} \right) \frac{ \partial_x \beta}{\beta} \right].
\end{align}
With what we refer to as the symmetric and antisymmetric functions $H(\theta,\mu)$ and $h(\theta,\mu)$, respectively, defined as
\begin{eqnarray}
    H(\theta,\mu) = 2\pi (\dr 1(\theta))(\dr \varphi(\theta,\mu))^2 (\dr 1(\mu))  \frac{\left( v^{\rm eff}(\theta) - v^{\rm eff}(\mu) \right)^4}{|v^{\rm eff}(\theta) - v^{\rm eff}(\mu)| },  \\ 
    h(\theta,\mu) = 2\pi (\dr 1(\theta))(\dr \varphi(\theta,\mu))^2 (\dr 1(\mu))  \frac{\left( v^{\rm eff}(\theta) - v^{\rm eff}(\mu) \right)^3}{|v^{\rm eff}(\theta) - v^{\rm eff}(\mu)| } .
\end{eqnarray}
Note that both of these functions are invariant under Galilean boosts, in addition to the symmetric/ antisymmetric behaviour under interchange of the arguments $H(\theta,\mu) = H(\mu, \theta)$ and $h(\theta,\mu) = - h(\mu,\theta)$. Note that the $\partial_x \beta$ vanishes, due to similar considerations when considering the energy diffusion in sec.~\ref{app:energy_hd_eq}. 

Combined, these considerations allow the momentum diffusion contribution to be written as
\begin{eqnarray}
     \frac 1 2  \partial_x \left( \int d\theta  v^{\rm eff}(\theta) (\dr 1(\theta))^2 \left[\tilde{ D} \partial_x n\right](\theta)\right) &=&\frac 1 {4} \int d\theta d\mu \, \partial_x \left[ \frac{\partial_\theta n}{\partial_\theta \epsilon} \frac{\partial_\mu n}{\partial_\mu \epsilon} H(\theta,\mu) \frac{\partial_x \eta}{\beta}\right]  .
      \end{eqnarray}
If the integral is evaluated in the $1/\beta \to 0$ limit then it follows that 
\begin{eqnarray}
 \frac 1 2    \partial_x \left( \int d\theta  v^{\rm eff}(\theta) (\dr 1(\theta))^2 \left[\tilde{ D} \partial_x n\right](\theta)\right) 
    &=&\frac {8\pi} {\beta} \partial_x \left[ v_F (\dr \varphi(q,-q))^2 (\partial_x \eta)\right],
\end{eqnarray}
and the symmetric function at the Fermi points reduces to
\begin{eqnarray}
    H(q,-q) = H(-q,q) = 2\pi (\dr \varphi(q,-q))^2 (\dr 1(q))^2 8 v_F^3.
\end{eqnarray}
Equivalently, since the derivative and zero temperature limits commute, see App.~\ref{app:commutativity}, the derivative of the integrand can be taken prior to integration. This procedure is carried out explicitly in App.~\ref{app:spatial_drv_preamble} and demonstrated as equivalent. 

\subsection{Evaluation of derivative of the momentum diffusion kernel}
\label{app:spatial_drv_preamble}
Here the spatial derivative of the integrand is considered in detail and proven to coincide with the reported visosity $\mu(\rho,\beta)$. To evaluate the spatial derivative of the integrand there are two key identities. The first being a result of the low-temperature behaviour of the Fermi sea and the second taking into account the melting on the edges of the filling factor
\begin{eqnarray}
    \partial_\theta n &\to& - \sum_{\sigma = \pm} \sigma \delta (\theta - \theta^\sigma),\\
    \label{eq:niceId}
   \beta (2n  - 1) \partial_\theta n &\to&  - \sum_{\sigma = \pm} \frac{\sigma}{(\partial_\theta \epsilon)(\theta^\sigma)} \delta' (\theta - \theta^\sigma),
\end{eqnarray}
the second line being proven in App.~\ref{app:niceId_deriv}. These relations then yield a particularly useful combined expression
\begin{eqnarray}
\label{eq:combined_id}
   \partial_x \left( \frac{\partial_\theta n}{\partial_\theta \epsilon}\right)
   = - \sum_{\sigma =\pm} \frac{\sigma}{(\partial_\theta \epsilon)(\theta^\sigma)} \delta'(\theta-\theta^\sigma) \frac{\partial_x\epsilon}{\partial_\theta \epsilon}. 
\end{eqnarray}
Recall that $q = \frac{\theta^+ - \theta^-}{2}$, such that $\theta^\pm = \frac{ \theta^+ + \theta^-}{2} \pm q$. So acting on each term with the spatial derivative yields
      \begin{eqnarray}
&&      \partial_x \left( \int d\theta  v^{\rm eff}(\theta) (\dr 1(\theta))^2 \left[\tilde{ D} \partial_x n\right](\theta)\right)      = \frac{1}{2}  \int d\theta \int d\mu \sum_{\sigma,\sigma'}   \frac{\sigma' \delta(\mu - \theta^{\sigma'})}{\partial_\mu \epsilon} \frac{\sigma \delta (\theta- \theta^\sigma)}{\partial_\theta \epsilon}\partial_x \Big(\frac{(\partial_x \eta)   H(\theta,\mu)}{\beta} \Big), \nn\\
      &&-  \int \frac{d\theta \int d\mu}{2\beta} \sum_{\sigma,\sigma'}   \frac{\sigma' \delta(\mu - \theta^{\sigma'})}{\partial_\mu \epsilon(\theta^{\sigma'})} \frac{\sigma \delta (\theta- \theta^\sigma)}{\partial_\theta \epsilon(\theta^\sigma)}  \left[ \partial_\theta \left(\frac{\partial_x \epsilon(\theta)}{\partial_\theta \epsilon} H(\theta,\mu) \right)+ \partial_\mu \left( \frac{\partial_x \epsilon(\mu)}{\partial_\mu \epsilon} H(\theta,\mu) \right) \right] (\partial_x \eta).
\end{eqnarray}
The second term picking up an overall `$-$' due to an integration by parts after inserting the appropriate identities. The first term will be considered in more detail later, but for now a lot of effort is needed to determine a more reasonable form of the second term.


By combining many of the dressed function identities and noting that the symmetric function $H(\theta,\mu)$ is invariant under interchange of $\theta,\mu$ it is possible to write identities for the derivatives of the symmetric function as
\begin{eqnarray}
    (\partial_\theta + \partial_\mu) H(\theta,\mu)\bigg|_{\theta,\mu = q,-q} &=& 0,\nn\\
    \partial_q H(\theta,\mu)\bigg|_{\theta,\mu = q,-q} &=& H(q,-q) \left( 6 \dr \varphi(q,q) - 4 \dr \varphi(q,-q) \right), \nn\\  
    (\partial_\theta - \partial_\mu) H(\theta,\mu)\bigg|_{\theta,\mu = q,-q} &=& \frac{H(q,-q)}{v_F}(3-4 v_F \dr \varphi(q,-q) - 2 v_F \dr \varphi(q,q)).
\end{eqnarray}

By combining these derivative identities for the symmetric function and velocity the low-temperature diffusion can be evaluated to be
 \begin{eqnarray}
 &\partial_x& \left( \int d\theta  v^{\rm eff}(\theta) (\dr 1(\theta))^2 [\tilde{ D} \partial_x n](\theta)\right)  = \frac{ 2 H(q,-q) }{(\partial_\theta \epsilon(\theta^+))^2} \partial_x \left( \frac{\partial_x \eta}{\beta} \right)  + \frac{ 2 (\partial_x \eta)( \partial_x q)}{\beta (\partial_\theta \epsilon(\theta^+))^2} \left(\partial_q  + (\partial_\theta - \partial_\mu)  - \frac{2\partial_\theta v^{\rm eff}(\theta)}{v^{\rm eff}(\theta)}  \right) H(\theta,\mu) \bigg|_{\theta,\mu=q,-q} , \nn\\
 &&= \frac{  \partial_x^2 \eta}{\beta(\partial_\theta \epsilon(\theta^+))^2} H(q,-q) + \frac{   (\partial_x \eta)( \partial_x q)}{\beta v_F (\partial_\theta \epsilon(\theta^+))^2} H(q,-q) \left[1+ 4 v_F \dr \varphi(q,q) - 4 v_F \dr \varphi(q,-q) \right].
\end{eqnarray}
This formula can be rewritten more succinctly as 
\begin{eqnarray}
    \partial_x \left( \int d\theta  v^{\rm eff}(\theta) (\dr 1(\theta))^2 \left[\tilde{ D} \partial_x n\right](\theta)\right)    &=&16\pi\partial_x \left[ \frac {v_F}{\beta} (\dr \varphi(q,-q))^2 (\partial_x \eta)\right].
\end{eqnarray}
With the viscosity $\mu(q,\beta)$ identified as
\begin{eqnarray}
\label{eq:appviscous_def}
    \mu(q,\beta) = \frac{8 \pi}{\beta} v_F ( \dr \varphi(q,-q))^2.
\end{eqnarray}

\section{Equation for the energy density}
\label{app:energy_hd_eq}
At the Euler scale, the low-temperature energy evolution equation was found to be a restatement of the density and momentum relations. By adding diffusion this is no longer the case, due to the presence of the temperature field in the density and momentum equations. Since the diffusion is an $O(T)$ effect, there is no avoiding the temperature field. A standard calculation akin to that done in the Euler case, but now retaining the diffusive effects, then implies
\begin{eqnarray}
\frac 1 \rho \left(\partial_t E + \partial_x J_E\right) 
&=& \partial_t e +  \frac {\mathcal P_s} \rho    \partial_x \eta + \eta \partial_x e + 0 + \frac{\eta}{\rho} D(p) = \frac{1}\rho D(E).
\end{eqnarray}
To close our diffusive equations at low-temperature there are several steps. First, the diffusive term is evaluated, as it was for the momentum. With the low-temperature energy diffusion characterized, the energy field can be recast in terms of a temperature field, which closes the set of equations. The density, momentum, and temperature evolution equations are then brought together and collectively understood as constituting low-temperature viscous GHD.

The diffusive correction to the energy expression is then computed here, where for convenience of notation we introduce $v_E = \frac{\dr{( \theta_s^2)}}{2(\dr 1)}$ with the note that
\begin{eqnarray}
  D(E) &=&\partial_x \int \frac{d\theta d\mu}{2}  \frac{\dr{(\theta^2)}(\theta)}{2} (\dr 1(\theta)) \tilde D(\theta,\mu) \partial_x n(\mu) = \partial_x \int \frac{d\theta d\mu}{2} \left( v_E(\theta) +  \eta v^{\rm eff}(\theta) + \frac{\eta^2}{2}\right) (\dr 1(\theta))^2 \tilde D(\theta,\mu) \partial_x n(\mu),\nn\\
  \label{eq:sm_energyDiffusionShift}
  &=&D_{v_E} + \mu(\rho,\beta) (\partial_x \eta)^2 + \eta D(p) + \eta^2 D(\rho).
\end{eqnarray}
Only the first term $D_{v_E}$ is new, the viscosity in the second term has been calculated already, see~(\ref{eq:appviscous_def}), the third term is exactly the diffusive correction to momentum, and the third term the diffusive correction to density, which vanishes. Recall now the definition of the anti-symmetric function $h(\theta,\mu)$ 
\begin{eqnarray}
    h(\theta,\mu) =  2\pi \frac{(\dr 1(\theta)\dr 1(\mu)) (\dr \varphi(\theta,\mu))^2}{\beta}  \frac{(v^{\rm eff}(\theta) - v^{\rm eff}(\mu))^3}{|v^{\rm eff}(\theta) - v^{\rm eff}(\mu)|},
\end{eqnarray}
which is helpful for simplifying the calculation.
With the antisymmetric function and noting that $v_E(q) = v_E(-q)$ the diffusive contribution $D_{v_E}$, after applying the spatial derivative, is given by 
\begin{eqnarray}
    D_{v_E}&=&\int \frac{d\theta d\mu}{4\beta }\left( \frac{\partial_\theta n}{\partial_\theta \epsilon} \frac{\partial_\mu n}{\partial_\mu \epsilon} \partial_x(h(\theta,\mu) (v_E(\theta) - v_E(\mu)) \partial_x \eta)  +\right. \nn\\
    &&+\frac{\partial_\theta n }{\partial_\theta \epsilon} \frac{\partial_\mu n }{\partial_\mu \epsilon}(\partial_x q) (\partial_x \eta)\left(\partial_\theta \frac{h(\theta,\mu)(v_E(\theta) - v_E(\mu))}{v^{\rm eff}(\theta)}  + \partial_\mu\frac{h(\theta,\mu)(v_E(\theta) - v_E(\mu))}{v^{\rm eff}(\mu)} \right), \nn\\
    &&\left. + \frac{\partial_\theta n }{\partial_\theta \epsilon} \frac{\partial_\mu n }{\partial_\mu \epsilon}(\partial_x \eta)^2  \left[\partial_\theta [h(\theta,\mu)(v_E(\theta) - v_E(\mu))] + \partial_\mu[h(\theta,\mu)(v_E(\theta) - v_E(\mu)] \right]\right).
\end{eqnarray}
Note that we have dropped the terms proportional to $\partial_x \beta$, since these terms contain $O((v_E(\theta) - v_E(\mu))^2)$ and so will always vanish when $\theta,\mu \to q,-q$.
The other terms can be determined as
\begin{eqnarray}
    \int \frac{d\theta d\mu}{4 \beta}\frac{\partial_\theta n }{\partial_\theta \epsilon} \frac{\partial_\mu n }{\partial_\mu \epsilon}\left(\partial_\theta \frac{h(\theta,\mu)(v_E(\theta) - v_E(\mu))}{v^{\rm eff}(\theta)}  + \partial_\mu\frac{h(\theta,\mu)(v_E(\theta) - v_E(\mu))}{v^{\rm eff}(\mu)} \right) = 0.
\end{eqnarray}
Along with the second term contributing 
\begin{eqnarray}
\label{eq:energyDiff1}
    \int \frac{d\theta d\mu}{4 \beta} \frac{\partial_\theta n }{\partial_\theta \epsilon} \frac{\partial_\mu n }{\partial_\mu \epsilon}(\partial_x \eta)^2  \left[(\partial_\theta + \partial_\mu)[h(\theta,\mu)(v_E(\theta) - v_E(\mu)] \right] = \frac{16 \pi v_F }{\beta} (\dr \varphi(q,-q))^2 (\partial_x \eta)^2 .\nn\\
\end{eqnarray}
Likewise, the only surviving part of the first term is the spatial derivative of $v_E$. By use of dressed function identities it then follows that
the first term can be written as
\begin{eqnarray}
\label{eq:energyDiff2}
    \int \frac{d\theta d\mu}{4\beta } \frac{\partial_\theta n}{\partial_\theta \epsilon} \frac{\partial_\mu n}{\partial_\mu \epsilon} \partial_x(h(\theta,\mu) (v_E(\theta) - v_E(\mu)) \partial_x \eta) = -\frac{16 \pi v_F}{\beta} (\dr \varphi(q,-q))^2 (\partial_x \eta)^2.
\end{eqnarray}
Taken together Eqs.~\eqref{eq:energyDiff1} and \eqref{eq:energyDiff2} imply the diffusive contribution from $D_{v_E}$ vanishes
\begin{eqnarray}
D_{v_E} =    \partial_x \int d\theta d\mu \, v_E(\theta) (\dr 1(\theta))^2 \tilde D(\theta,\mu) \partial_x n(\mu) = 0.
\end{eqnarray}
Consequently, the total energy diffusion is given by
\begin{eqnarray}
    D(E) &=& D_{v_E} + (\partial_x \eta)^2  \left( \frac{8 \pi}{\beta} (\dr \varphi(q,-q))^2 v^{\rm eff}(q) \right) + \eta D(p) + \eta^2 D_{\rho},\nn\\
    &=& (\partial_x \eta)^2  \left( \frac{8 \pi}{\beta} (\dr \varphi(q,-q))^2 v^{\rm eff}(q) \right) + \eta D(p).
\end{eqnarray}
Finally, this is combined with the energy evolution equation, which after some simplification reads
\begin{eqnarray}
\partial_t e + \eta \partial_x e + \frac{\mathcal{P}_s}{\rho} \partial_x \eta    &=& \frac{\mu}{\rho} (\partial_x \eta)^2  
\end{eqnarray}

\subsection{Thermal Conductivity }
{

First consider the form of the GHD equation (restricted to two Fermi points). As we have already seen, the hydrodynamic equations can be written in the boosted frame.  
So here Eq.~\eqref{eq:sm_energyDiffusionShift} is replicated for convenience
\begin{eqnarray}
\partial_t e + \eta \partial_x e+\frac{\mathcal P_s}{\rho} \partial_x \eta  = \frac{D_{v_E} + \mu (\partial_x \eta)^2}{\rho},
\end{eqnarray}
with the reminder that $e = E/\rho$. The thermal conductivity will be proportional to $\partial_x \beta$ type terms, hence must come from a higher order temperature correction to the boosted diffusion $D_{v_E}$. The explicit form of this diffusion kernel is recalled as
\begin{eqnarray}
     D_{v_E} \sim \partial_x  \left(\int  d\theta  d\alpha\,\frac{\pi \dr{(\theta^2_s)}(\theta)}{2\beta^2 \dr 1(\theta)}   \frac{\partial_\theta n}{\partial_\theta \epsilon} \frac{\partial_\alpha n}{\partial_\alpha \epsilon}  \frac{(v(\theta) - v(\alpha))^2}{|v(\theta) - v(\alpha)|}  \dr 1(\theta) \dr 1(\alpha) (\dr \varphi(\theta,\alpha))^2\left( \frac{\partial_x (\beta\epsilon(\alpha))}{\dr 1(\alpha)} - \frac{\partial_x(\beta \epsilon(\theta))}{\dr 1(\theta)} \right)\right)  .\nn\\
\end{eqnarray}
To determine the low temperature thermal conductivity $\kappa(\rho,T)$ we note that the term must appear as a higher order temperature correction to the energy diffusion, hence will be the first surviving term proportional to $\partial_x \beta$. With this in mind it is furthermore convenient to note that the low temperature thermal energy is given by
\begin{eqnarray}
    \frac{\epsilon(\theta)}{\dr 1(\theta)} \sim  \frac{\dr{(\theta^2)}(\theta)}{2 \, \dr 1(\theta)} - \mu .
\end{eqnarray}
With the additional note that $\mu$ is the local chemical potential, and hence constant so that it vanishes in the difference. This relation allows us to rewrite the part of the energy diffusion kernel containing the thermal conductivity as
\begin{eqnarray}
     D_{v_E} \overset{{O(\partial_x \beta)}}{\sim} \partial_x \left(\frac{\pi\partial_x \beta}{2\beta^2 } \int  d\theta  d\alpha\,   \frac{\partial_\theta n}{\partial_\theta \epsilon} \frac{\partial_\alpha n}{\partial_\alpha \epsilon}  \frac{(v(\theta) - v(\alpha))^2}{|v(\theta) - v(\alpha)|}  \dr 1(\theta) \dr 1(\alpha) (\dr \varphi(\theta,\alpha))^2\left( \frac{\epsilon(\alpha)}{\dr 1(\alpha)} - \frac{ \epsilon(\theta)}{\dr 1(\theta)} \right)^2 \right) 
\end{eqnarray}
Similarly to what we've previously encountered the relevant contributions are those at the Fermi points, which vanish at leading order as can be checked by noting. The corrections are implemented by $\dr{f}(q_T) = \dr{f}(q) + \frac{\delta \dr{f}}{\delta q_T} \bigg|_{q} (q_T - q)$, along with both $\epsilon(q_T) = 0$ and the small temperature Fermi point corrections found in Eq.~\eqref{eq:fermi_point_shift} and reproduced here as
\begin{eqnarray}
     2 (q_T - q )= - \frac{\pi^2 T^2}{3 \dr \theta v_F } \left( 2  V(q) + \left[ \frac 1 {v_F(q)} - 2U(q)\right]   \right),
\end{eqnarray}
where $V(\theta) = \dr \varphi(\theta,q) - \dr \varphi(\theta,-q)$ and $U(\theta) = \dr \varphi(\theta,q)+\dr \varphi(\theta,-q)$.  The variation of $\epsilon/ \dr 1$ is readily determined from the zero temperature identity in Eq.~\eqref{eq:zero_temp_fluct} as being
\begin{eqnarray}
\frac{\delta}{\delta q} \left( \frac{\epsilon(q)}{\dr 1(q)}  \right) =  \frac{\delta v_E}{\delta q} \bigg|_{q} (q_T - q) = v(q) (q_T - q).
\end{eqnarray}
Combined with the Fermi point expression the
\begin{eqnarray}
    D_{v_E} 
    \overset{O(\partial_x T)}{\sim} \partial_x \left[\frac{4 v_F \pi^5  T^4  \partial_x T}{9 (\dr \theta v_F)^2 } (\dr \varphi(q,-q))^2   \left( \frac{1 - 4 v_F \dr\varphi(q,-q)}{2 v_F} \right)^2\right].
\end{eqnarray}
To cast this in terms of the Luttinger liquid parameter $K = (\dr 1(q))^2 $, the identities $2 K \partial_\rho v_F = (1 - 4 \dr \varphi(q,-q) v_F)$ and the expression for viscosity $\mu(\rho,T)$ from Eq.~\eqref{eq:app_visc_final} are utilized. It follows that the thermal conductivity is given by
\begin{eqnarray}
        \kappa(\rho,\eta,T) \partial_x T = \frac{4 \pi^5 T^4 \partial_x T}{9} v_F K^3 \left( \frac{(\partial_\rho v_F) (\partial_\rho  K)}{\rho v_F} \right)^2
\end{eqnarray}
In the notation with $\rho = \frac{1}{2\pi} \int d\theta \, \dr 1(\theta) n $ this expression is equivalently given by
\begin{eqnarray}
        \kappa(\rho,\eta,T) \partial_x T  
        = \frac{T^4 \partial_x T}{32} \frac{K \mu(\rho,T)}{T} \left( \partial_\rho (\tilde \chi_e K) \right)^2 .   
\end{eqnarray}
This can also be written in terms of $\gamma$ finding that 
\begin{eqnarray}
    \frac{\kappa(\gamma,\eta,T)}{c} \frac{\partial_x T}{c^2} = \frac{\tau^3  \mu(\gamma,\tau) K \gamma^4}{32}   [\partial_\gamma (\tilde \chi_e(\gamma) K) ]^2 \, \partial_x \tau.
\end{eqnarray}
With additional simplifications in the final expression involving the observation that $\tilde \chi_e K = \frac{\pi^2}{3 v_F^2}$
This thermal conductivity is now combined with the rest of the hydrodynamic results to find 
\begin{eqnarray}
\left( \partial_t T + \eta \partial_x T\right) = - \frac{\mathcal P_s}{\rho T \chi_e} \partial_x \eta +  (\partial_x \eta)^2 \frac{ \mu+ O(T^3)}{\rho T \tilde \chi_e}  + \frac{\partial_x(\kappa \partial_x T) }{\rho T \tilde \chi_e}.
\end{eqnarray}
It should be noted that the thermal conductivity is subleading, $\kappa \sim O(T^4)$, hence competes with other subleading corrections to the viscosity, pressure, and energy susceptibility. As a consequence, although physically significant, the thermal conductivity is not expected to be relevant in the low temperature limit. 
}

\section{Dynamics of the temperature and inclusion of other conserved quantities}
\label{app:additional_Lagrange}

Our goal here is to compute the evolution of the temperature and also to consider the impact of additional Lagrange parameters (temperatures) that are allowed to evolve at low temperatures in the integrable gas. This amounts to redefining the thermal energy and so obtain a more general local filling function. The simplest example is the case of two additional conserved quantities, $Q_3$ and $Q_4$
\begin{eqnarray}
    \beta \epsilon = \beta \left( \frac{(\theta-\eta)^2}{2} - \mu + \gamma_3  \frac{(\theta-\eta)^3}{3!} +  \gamma_4  \frac{(\theta-\eta)^4}{4!} \right) + \int d\alpha \, \varphi(\theta-\alpha) \ln \left| 1 + \e^{-\beta \epsilon} \right|.
\end{eqnarray}


To begin only those temperatures that lead to an even Fermi weight will be considered, which will then be generalized to the anti-symmetric case.

Consider the general hydrodynamic charges corresponding to these generalized temperatures, which are explicitly given by
\begin{align}
    &\frac{\langle Q_2 \rangle}{\rho} = e - \frac{\eta^2}{2}  \, ,& \rho \langle e \rangle = \int d\theta \frac{(\theta^2)^{\rm dr}}{2} n(\theta)\, , \nn\\
     &\frac{ \langle Q_3 \rangle }{ \rho} = \langle q_3 \rangle-  \eta \langle q_2 \rangle  - \frac{\eta^3}{3!} \, , & \rho \langle q_3 \rangle  = \int d\theta \frac{(\theta^3)^{\rm dr}}{3!} n(\theta) \, , \nn\\
    &\frac{ \langle Q_4 \rangle}{\rho} = \langle q_4 \rangle  -  {\eta} \langle q_3 \rangle - \frac{\eta^2}{2} \langle q_2 \rangle  - \frac{\eta^4}{4!} \, , & \rho \langle q_4 \rangle = \int d\theta \frac{(\theta^4)^{\rm dr}}{4!}  n(\theta)\, .
\end{align}
Likewise, the corresponding diffusive terms are characterized by generalized viscosities $\mu_n$, given by the low-temperature diffusion for an arbitrary conserved charge. At leading order in temperature the general form is given by 
\begin{eqnarray}
\label{eq:generalViscosity}
  \mu_n &=&\frac{8 \pi}{\beta} \frac{ \dr q_n(q)}{\dr 1(q)} ( \dr \varphi(q,-q))^2 = \frac{ 8 \pi v_n(q)}{\beta} ( \dr \varphi(q,-q))^2 \sim O(T),
\end{eqnarray}
where $v_n = \dr q_n / \dr 1$ and noting that the only nontrivial viscosity terms are odd. The set of hydrodynamic equations for symmetric Fermi weights with an additional charge is given by 
\begin{eqnarray}
    \partial_t \rho + \partial_x (\eta \rho) &=& 0,\nn\\
    \partial_t\eta + \frac{1}{2} \partial_x ( \eta^2) +\frac{\partial_x  \mathcal{P}_s(\rho,T) }{\rho} &=&  
\frac{1}{\rho} \partial_x (\mu (\rho,T) \partial_x \eta )\, ,\nn \\
\label{eq:evolutioncharges}
    \partial_t \langle q_n \rangle + \eta \partial_x  \langle q_n \rangle + \frac{\mathcal{P}_{n}}{\rho} \partial_x \eta  &=& \frac{\mu_{n-1} }{\rho} (\partial_x \eta)^2 ,
\end{eqnarray}
where the generalised static pressure given by $\mathcal{P}_n = \int d\theta \, \dr \theta n(\theta) q_n$, not to be confused with the stationary pressure $\mathcal{P}_s$, and the generalized viscosity $\mu_n$ is defined in eq.~\eqref{eq:generalViscosity}.  We now move to the evolution of the temperatures, 

\begin{eqnarray}\label{eq:tempeq1}
    \frac{\delta \langle q_n \rangle}{\delta T} (\partial_t T + \eta \partial_x T) + \sum_{i=3}^4 \frac{\delta \langle q_n \rangle}{\delta \gamma_i} (\partial_t \gamma_i + \eta \partial_x \gamma_i) =  P^T_n  \partial_x \eta + \frac{\mu_n}{\rho} (\partial_x \eta)^2,
\end{eqnarray}
along with the identification 
\begin{eqnarray}
\label{eq:lowT_prod}
 P^T_n = \Big[ \rho   \partial_\rho \langle q_n \rangle - \frac{\mathcal{P}_{n-1}}{\rho} \Big] - \Big[  \rho  \partial_\rho \langle q_n \rangle - \frac{\mathcal{P}_{n-1}}{\rho} \Big]_{T=0}. 
\end{eqnarray}
The general correction $P_n^T$ is given in eq.~\eqref{eq:pressureCorrection}.
This expression then determines the evolution of the temperature, eq. \eqref{eq:tempeq1} in a thermal state, using $\delta e/\delta T = T \tilde{\chi}_e$, 
\begin{eqnarray}
    ( \partial_t T + \eta \partial_x T ) =     \frac{T P_2^T}{ \tilde{\chi}_e}  \partial_x \eta + \frac{\mu(\rho,T)}{\rho T  \tilde \chi^T_e } (\partial_x \eta)^2
\end{eqnarray}
Let us now consider the case where also different temperatures can be activated. These additional Lagrange parameters satisfy the general general relations at low temperatures of
\begin{equation}
    \delta \langle q_n \rangle / \delta T \sim T  \quad ,  \delta \langle q_n \rangle / \delta \gamma_i \sim T^2, 
\end{equation}
which follows from the general expressions for the low temperature charges from eq.~\eqref{eq:modifiedCharge}.
Recombining this with the expression for the evolution of the Lagrange parameters, for both $q_2$ and $q_4$, with $\chi^{(n)}_{\gamma_j} =  \frac{1}{T^2}\frac{\partial \langle q_n \rangle}{\partial \gamma_j}$, and $\chi^{(n)}_T = \frac{1}{T}\frac{\partial \langle q_n \rangle}{\partial T}$ find the evolution equations 
\begin{eqnarray}     
T \chi_T^{(2)} (\partial_t T + \eta \partial_x T) + \sum_{i=3}^4 T^2 \chi_{\gamma_j}^{(2)} (\partial_t \gamma_i + \eta \partial_x \gamma_i) =   P_2^T  \partial_x \eta + \frac{\mu_1}{\rho} (\partial_x \eta)^2, \nn\\
T \chi_T^{(4)} (\partial_t T + \eta \partial_x T) + \sum_{i=3}^4 T^2 \chi_{\gamma_j}^{(4)} (\partial_t \gamma_i + \eta \partial_x \gamma_i) =   P_4^T  \partial_x \eta + \frac{\mu_3}{\rho} (\partial_x \eta)^2.
\end{eqnarray}
Simple algebra then implies the following evolution equation for the additional Lagrange parameters
\begin{eqnarray}
    \sum_{i=3}^4 T \left(\frac{\chi_{\gamma_j}^{(4)}}{\chi_{T}^{(4)}} -\frac{\chi_{\gamma_j}^{(2)}}{\chi_{T}^{(2)}} \right) (\partial_t \gamma_i + \eta \partial_x \gamma_i) =  \frac 1 T \left( \frac{P_4^T}{{\chi_{T}^{(4)}}} - \frac{P_2^T}{{\chi_{T}^{(2)}}} \right)  \partial_x \eta + \left( \frac{\mu_3}{T \chi_{T}^{(4)}\rho} -\frac{\mu_1}{T \chi_{T}^{(2)}\rho} \right) (\partial_x \eta)^2 .
\end{eqnarray}
This appears to be $O(T^{-1})$, however, by inserting the leading order terms for $\mu_1$, $\mu_3$, and $\chi_T^{(n)}$ then the diffusive part vanishes
\begin{eqnarray}
    \left( \frac{\mu_3}{T \chi_{T}^{(4)}\rho} -\frac{\mu_1}{T \chi_{T}^{(2)}\rho} \right)  =  O(T^2).
\end{eqnarray}
With the note that the $O(T^2)$ is due to higher order corrections to the viscosity and susceptibilities. In the case with only a single additional Lagrange parameter, remove $\gamma_3$, the evolution equation for $\gamma_4$ is then written as
\begin{eqnarray}
    \partial_t \gamma_4 + \eta \partial_x \gamma_4 =  \frac 1 {T^2} \left( \frac{ P^T_4 \chi^{(2)}_T- P_2^T \chi^{(4)}_T}{\chi_T^{(4)} \chi_{\gamma_4}^{(2)} - \chi_{\gamma_4}^{(4)} \chi_T^{(2)}  } \right)  \partial_x \eta \sim O(T^0),
\end{eqnarray}
this can be reinserted to determine the temperature evolution as
\begin{eqnarray}
    \partial_t T + \eta \partial_x T &=& \frac 1 T \frac{P^T_2  \partial_x \eta}{\chi_T^{(2)}} + \frac{ \mu_1 (\partial_x \eta)^2}{T \chi_T^{(2)}} + \frac 1 T \frac{ \chi^{(2)}_{\gamma_4}}{\chi_T^{(2)}} \left( \frac{P_2^T \chi^{(4)}_T - P^T_4 \chi^{(2)}_T}{\chi_T^{(4)} \chi_{\gamma_4}^{(2)} - \chi_{\gamma_4}^{(4)} \chi_T^{(2)}  } \right) \rho \partial_x \eta. \nn\\
\end{eqnarray}
Note that inserting $P_n^T \sim O(T^2)$ then in the absence of diffusion ($\mu_j \to 0$) the growth for both $\gamma_3$ and $\gamma_4$ are $O(T)$, whereas $T$ changes as $O(T^0)$.
\\

A similar analysis is now carried out for the case of a non-symmetric Fermi weight. Here, although the hydrodynamic evolution equations are more complicated, the general expressions are
\begin{eqnarray}
   0&=& \partial_t \rho + \partial_x (\rho \eta) , \nn\\
   \label{eq:generalizedEvolutionEq}
       0&=& \partial_t \langle q_k \rangle + \eta \partial_x \langle q_k \rangle - \frac{\langle q_{k-1}\rangle}{\rho} \partial_x \mathcal{P}_1 + \frac 1 \rho \partial_x \mathcal P_k +\frac{\mathcal{P}_{k-1}}{\rho} \partial_x \eta,
\end{eqnarray}
such that the evolution of the Lagrange parameters are given in general by
\begin{eqnarray}
    T (\partial_t T + \eta \partial_x T) \chi^{(n)}_T + \sum_{j=3} (\partial_t \gamma_j + \eta \partial_x \gamma_j) T^2 \chi_{\gamma_j}^{(n)} &=& \frac{\langle q_{n-1}\rangle}{\rho} \partial_x \mathcal{P}_1 - \frac{\partial_x \mathcal{P}_{n}}{\rho} + P_{n-1}^T  \partial_x \eta ,
\end{eqnarray}
where, as before, $P^T_{n-1} = \partial_\rho \langle q_n \rangle - \frac{\mathcal{P}_{n-1}}{\rho}$
It should be noted that $\langle q_1 \rangle =0$ and $ n \geq 2 $, furthermore, naively there appears to be $O(T^0)$ terms present on the RHS that would lead to $O(T^{-1})$ corrections overall. The $T=0$ identities imply, however, that $\langle q_{n-1} \rangle \partial_x \mathcal{P}_1/ \rho \sim v_n \partial_x q + O(T^2)$ and $\partial_x \mathcal{P}_n/\rho \sim v_n \partial_x q + O(T^2)$, so this apparent leading term vanishes. The corrections stemming from $O(T^2)$ corrections are combined into the shorthand variable $G_n \sim O(T^2)$ defined as
\begin{eqnarray}
     G_n  &=& P_{n-1}^T  \partial_x \eta+ \frac{\langle q_{n-1}\rangle}{\rho} \partial_x \mathcal{P}_1 - \frac{\partial_x \mathcal{P}_{n}}{\rho} \sim O(T^2) .
\end{eqnarray}
With these consideration the impact of the third and fourth additional Lagrange parameters $\gamma_3$ and $\gamma_4$ can be considered. Introduce $\widehat{\mathcal D} = \partial_t + \eta \partial_x$ and so determine that the evolution equations read
\begin{eqnarray}
    \widehat {\mathcal{D}}T &=&
     \frac{1}{T}\frac{G_2 \chi_T^{(3)} - G_3 \chi_T^{(2)} }{\chi_{\gamma_3}^{(3)} \chi_T^{(2)} - \chi_{\gamma_3}^{(2)} \chi_T^{(3)}} + \left(\frac{\chi_{\gamma_4}^{(3)} \chi_{\gamma_3}^{(2)} - \chi_{\gamma_4}^{(2)} \chi_{\gamma_3}^{(3)}}{\chi_{\gamma_3}^{(3)} \chi_T^{(2)} - \chi_{\gamma_3}^{(2)} \chi_T^{(3)}}  \right) T \widehat {\mathcal{D}} \gamma_4, \nn\\
    \widehat{\mathcal{D}} \gamma_3 &=& \frac {1} {T^2} \frac{G_3 \chi_T^{(2)} - G_2 \chi_T^{(3)}}{\chi^{(3)}_{\gamma_3} \chi_T^{(2)} - \chi_{\gamma_3}^{(2)} \chi_T^{(3)}}  - \frac{\chi_{\gamma_4}^{(3)} \chi_T^{(2)} - \chi_{\gamma_4}^{(2)}\chi_T^{(3)}}{\chi^{(3)}_{\gamma_3} \chi_T^{(2)} - \chi_{\gamma_3}^{(2)} \chi_T^{(3)}}\widehat{\mathcal{D}} \gamma_4, \nn\\
    \widehat{\mathcal{D}} \gamma_4 
    &=& \frac{G_4(\chi^{(3)}_{\gamma_3} \chi_T^{(2)} - \chi_{\gamma_3}^{(2)} \chi_T^{(3)}) - (\chi^{(4)}_T - \chi_{\gamma_3}^{(4)})\chi_T^{(3)}G_2 + (\chi_T^{(4)}-\chi_{\gamma_3}^{(4)}) \chi^{(2)}_T G_3}{T^2 \chi_{\gamma_4}^{(4)} (\chi^{(3)}_{\gamma_3} \chi_T^{(2)} - \chi_{\gamma_3}^{(2)} \chi_T^{(3)})} \nn\\
    &&+ \frac{\chi_T^{(4)}(\chi^{(3)}_{\gamma_4} \chi_{\gamma_3}^{(2)} - \chi_{\gamma_4}^{(2)} \chi_{\gamma_4}^{(3)})-\chi_{\gamma_3}^{(4)}(\chi_{\gamma_4}^{(3)}\chi_T^{(2)} - \chi_{\gamma_4}^{(2)} \chi_T^{(3)})}{\chi_{\gamma_4}^{(4)}(\chi^{(3)}_{\gamma_3} \chi_T^{(2)} - \chi_{\gamma_3}^{(2)} \chi_T^{(3)})}.
\end{eqnarray}
With the note that this implies that $T$, $\gamma_3$, and $\gamma_4$ evolve with corrections of order $O(T)$, $O(T^0)$, and $O(T^0)$ respectively.


\subsection{Derivation of the evolution of charges, eq. \eqref{eq:generalizedEvolutionEq}}
To prove the general form of the evolution equations the idea is to apply the relevant shifts to the conserved charges, so that they are being considered in the symmetric basis. These shifts are expanded and so the evolution equation for each charge field can be written in terms of a sum over evolution equation of conserved charges in the symmetric basis. These sums are rearranged and noted to vanish term-wise, which yields the general form of the evolution equation whenever the Fermi weight is symmetric. The diffusive case is then considered, which requires the addition of an additional sum that is treated analogously to the Euler case. Finally, the symmetric condition for the Fermi weight is relaxed and the fully general expression is obtained.
\\

In general the hydrodynamic evolution equations can be written with the notation $\rho \langle q_n \rangle = \int d\theta \frac{\theta^n}{n!} n(\theta) (\dr 1)$ and $\rho \langle j_n \rangle = \mathcal{P}_{n} = \int d\theta \, \frac{\theta^n}{n!} (\dr \theta) n(\theta)$. The density and $\eta$ fields are identified as
\begin{eqnarray}
    \rho = \int d\theta \, \dr 1 n, \nn\\
    \rho \eta = \int d\theta \, \dr \theta n \theta,
\end{eqnarray}
this choice of $\eta$ fixes $\langle q_1 \rangle =0$. These expression are used to rewrite the continuity equation as
\begin{eqnarray}
    \partial_t \left( \int d\theta (\theta + \eta)^n n \dr 1 \right) &+&     \partial_x \left( \int d\theta (\theta + \eta)^n n \dr{( \theta +\eta)}\right) = \sum_{k=0}^n \frac{n^{\underline{k}}}{n!} \left(\partial_t \left( \rho \eta^{n-k} \langle q_k \rangle \right) +\partial_x \left[ \rho \eta^{n-k+1} \langle q_k \rangle +\rho \eta^{n-k} \langle j_k \rangle \right] \right), \nn\\
 &=& \sum_{k=0}^n \frac{n^{\underline{k}}}{n!} \left(  \eta^{n-k} \partial_t(\rho \langle q_k \rangle)  + (n-k) \langle q_k \rangle \eta^{n-k -1 } \rho \partial_t \eta  \right.+  \eta^{n-k} \partial_x( \rho \eta \langle q_k \rangle )  \nn\\
     \label{eq:generalSum}
    && + (n - k ) \rho \langle q_k\rangle \eta^{n-k} \partial_x \eta  +\left. \eta^{n-k} \partial_x ( \rho \langle j_k \rangle) +(n-k) \rho \langle j_k \rangle  \eta^{n-k-1}\partial_x \eta    \right),\nn\\
    &=& \frac{1}{n!} \left(\partial_t \rho + \partial_x (\eta \rho) \right) + \frac{n^{\underline{n+1}}}{n! \eta} \left(\langle q_{n}\rangle \rho (\partial_t \eta + \eta \partial_x \eta) + \rho \langle j_n \rangle \partial_x \eta  \right) \nn\\
     \label{eq:generalEvolutionPf}
    &&  +     \sum_{k=1}^n \frac{n^{\underline{k}}}{n!} \eta^{n-k} \left(   \partial_t(\rho \langle q_k \rangle)  + \partial_x ( \rho \eta \langle q_k \rangle) + \partial_x ( \rho \langle j_k \rangle ) + \langle q_{k-1} \rangle \rho (\partial_t \eta  + \eta \partial_x \eta) +\rho \langle j_{k-1}\rangle \partial_x \eta \right) \nn\\
\end{eqnarray}
with the dropping power $n^{\underline{k}} = n (n-1) \dots (n-k+1)$, such that $n^{\underline{n+1}} = 0$. These terms must vanish term by term and so yields the conservation equations for the general evolution equations, written with $\mathcal P_n = \rho \langle j_n \rangle$, that for the Euler case read
\begin{eqnarray}
   0&=& \partial_t \rho + \partial_x (\rho \eta) , \nn\\
   0&=& \partial_t \eta + \eta \partial_x \eta + \frac{\partial_x \mathcal{P}_1}{\rho}, \nn \\
       0&=& \partial_t \langle q_k \rangle + \eta \partial_x \langle q_k \rangle - \frac{\langle q_{k-1}\rangle}{\rho} \partial_x \mathcal{P}_1 + \frac 1 \rho \partial_x \mathcal P_k +\frac{\mathcal{P}_{k-1}}{\rho} \partial_x \eta.
\end{eqnarray}
 When the Fermi weight is symmetric all odd, $n = 2k -1$, charges $\langle q_{2k-1} \rangle = 0$ and all even, $n = 2k$, currents $\langle j_{2 k} \rangle = 0$ this reduces the general formula to the hydrodynamic equations for symmetric Fermi weights.
\\

Now the diffusive corrections can be inserted and treated in a closely analogous fashion. Insert the diffusive correction 
\begin{eqnarray}
\partial_t \left( \int d\theta (\theta + \eta)^n n \dr 1 \right) +     \partial_x \left( \int d\theta (\theta + \eta)^n n \dr{(\theta + \eta)}  \right) &=& \frac{1}{2} \partial_x \int d\theta d\alpha \, \frac{(\theta + \eta)^n}{n!} D(\theta,\alpha) \partial_x ( n \dr 1)(\alpha) \nn\\
    &=& \partial_x \int d\theta \, \frac{(\theta + \eta)^n}{n!} \hat{\mu}(\theta)\partial_x \eta,
\end{eqnarray}
where the $\int d\theta \frac{\theta^n}{n!} \hat \mu(\theta) \partial_x \eta = \mu_n \partial_x \eta$ is chosen in analogy with the low temperature result, although the precise form does not affect the proof. From this notation we identify $\partial_x (\mu_n \partial_x \eta)$ as the diffusion in the symmetric basis for the $n$th conserved charge 
\begin{eqnarray}
     \left( \sum_{k=0} \frac{n^{\underline k}}{n!} \partial_x ( \mu_k \eta^{n-k} \partial_x \eta) \right)&=&\sum_{k=0}^{n} \frac{n^{\underline{k}}}{n!}\eta^{n-k}\left(   \mu_{k-1} (\partial_x \eta)^2 +   \partial_x (\mu_{k} \partial_x \eta) \right) .
\end{eqnarray}
With the diffusive matched to the term of the Euler evolution equation with matching prefactor $n^{\underline{k}}$.
\\

\subsection{Explicit form of \eqref{eq:lowT_prod} }
These additional Lagrange parameters depend on the specific form under consideration, however certain general expressions are still possible. For a single Fermi sea the even hydrodynamic charges are given by 
\begin{eqnarray}
\label{eq:modifiedCharge}
   \rho \langle q_n\rangle (\rho,T) 
   &=& \rho \langle q_n\rangle(\rho,0) + \frac{\pi^2 T^2 \dr 1 }{3 (\epsilon'(q,\gamma_4))^2} \dr{\left( \frac{\theta_s^{n-1}}{(n-1)!}\right)} .
\end{eqnarray}
Likewise, at constant density and low-temperature the hydrodynamic currents
\begin{eqnarray}
   (\rho \langle j_n \rangle)(\rho,T) = \int d\theta \, v q_n \dr 1 n = \int d\theta \, \theta \dr q_n n.
\end{eqnarray}
These currents are related to pressure by the simple relation $\mathcal P_n = \rho \langle j_n \rangle$. The general identity at low temperature and a single Fermi sea for constant density $\rho$ the corrections are given by
\begin{eqnarray}
    \rho \langle j_n \rangle (\rho,T)  &=& \rho  \langle j_n \rangle(\rho,0) +    \frac{\pi^2 T^2}{3 (\epsilon'(q,\gamma_4))^2 } \left( \dr \theta_s(q) \dr{\left(\frac{\theta_s^{n-1}}{(n-1)!}\right)}+ (\dr \theta \dr 1)  v_n \left[ \frac 1 {v_F} + 2(V(q) - U(q)) +f(\gamma_4) \right] \right).\nn\\
\end{eqnarray}

For the generalized pressure energy relation note that
\begin{eqnarray}
    \partial_\rho \langle q_n \rangle &=& \frac 1 \rho \int d\theta \, \dr{\left(\frac{\theta^n}{n!}\right)} (\dr 1) \partial_\rho n(\theta)  - \frac{\langle q_n \rangle}{\rho},\nn\\
    \frac{\mathcal{P}_{n-1}}{\rho^2}  &=& \frac{1}{\rho^2} \int d\theta \, \dr \theta n(\theta) \dr{\left(\frac{\theta^{n-1}}{(n-1)!}\right)} = -\frac{1}{\rho^2} \int d\theta \,\left[  \dr{\left(\frac{\theta^n}{n!}\right)} n(\theta) \partial_\theta \dr \theta + \dr{\left(\frac{\theta^n}{n!}\right)} (\partial_\theta n) \dr \theta \right], \nn\\
    &=& - \frac{ \langle q_n \rangle}{\rho} - \frac{\int d\theta \, \dr{\left(\frac{\theta^n}{n!}\right)} (\partial_\theta n )\dr \theta}{\rho^2}.
\end{eqnarray}
By subtracting these two formula the useful identity for computing the difference of the charge and pressure is determined as
\begin{eqnarray}
\Big[ \rho   \partial_\rho q_n - \frac{\mathcal{P}_{n}}{\rho} \Big] &=& \frac{\int d\theta \,\frac{ \dr{(\theta^n)}}{n!} \partial_\theta n\left( \rho \dr\theta  +\dr 1  \frac{\partial_\rho (\beta \epsilon)}{\partial_\theta (\beta \epsilon)}  \right)}{\rho} = \left[     \rho \partial_\rho \langle  q_n  \rangle  - \frac{ \mathcal{P}_{n-1}}{\rho}\right]_{T=0} + P^T_n.
\end{eqnarray}
The identity $4 \dr \varphi(q,-q) v_F = (1 - 2 K(\rho) \partial_\rho v_F(\rho))$ is applied and we notate the generalized effective velocities as $v_0 = 1$, $v_1 = v_F$, and $v_n = \dr{(\theta^n)} /(n! (\dr 1))$. From these identifications the scenario of a single Fermi sea with arbitrarily many Lagrange parameters can be written out by noting $\epsilon' = \dr \theta \left( 1 + \sum_{j=3} \gamma_j \frac{v_{j-1}}{v_F} \right)$. With these considerations the $O(T^2)$ corrections when there are two and then an arbitrary number of Lagrange parameters are respectively given by
\begin{eqnarray}
P^T_n &=&\frac{ \pi^2 T^2}{3 v_F \rho} \left( \frac{\partial_\rho v_{n-1}}{v_F} - \frac{( v_{n-1}+ v_F v_{n-2})}{\rho v_F} - \frac{2 v_{n-1} \partial_\rho v_F}{v_F^2} - \frac{2 v_{n-1}}{v_F \rho} K(\rho) \partial_\rho v_F \right) , \nn\\
\label{eq:pressureCorrection}
P_n^T &=&  \frac{ \pi^2 T^2}{3 v_F \rho} \left[\frac{\left( \frac{\partial_\rho v_{n-1}}{v_F} - \frac{( v_{n-1}+ v_F v_{n-2})}{\rho v_F} - \frac{2 v_{n-1} \partial_\rho v_F}{v_F^2} - \frac{2 v_{n-1}}{v_F \rho} K(\rho) \partial_\rho v_F \right)}{\left(1 + \sum_{j=3} \gamma_j \frac{ v_{j-1}}{v_F}\right)^2} - 2\frac{v_{n-1}\sum_{j=3} \gamma_j \left[ \frac{\partial_\rho v_{j-1}}{v_F} - \frac{v_{j-1}\partial_\rho v_F}{v_F^2}  \right]}{\left( 1 + \sum_{j=3} \gamma_j \frac{v_{j-1}}{v_F} \right)^3}  \right]. \nn\\
\end{eqnarray}


\subsection{Small temperature corrections to non-symmetric Fermi weights}

For the case of symmetric Fermi weights the pressure and charge can be straightforwardly expanded at low temperature and constant density in the case of symmetric Fermi points as
\begin{eqnarray}
    \mathcal{P}_{n} &=&\mathcal{P}_{n}(\rho,0) + \frac{ \pi^2 T^2 \rho}{6 (\epsilon'(q))^2 } \left( v_{n-1}  + \frac{v_n}{v_F} 2   K(\rho) \partial_\rho v_F  \right),\nn\\
   \rho \langle q_n \rangle (\rho,T) &=& \rho \langle q_n \rangle(\rho,0) + \frac{\pi^2 T^2 \rho}{6 (\epsilon'(q))^2} \frac{ v_{n-1}}{v_F}.
\end{eqnarray}
The case of non-symmetric Fermi points is more complicated, however a similar picture holds. Specifically, whenever there is a single Fermi sea the state energy $\beta \epsilon$, for $k = x, t$ reads
\begin{eqnarray}
\partial_k(\beta \epsilon) = -(\partial_k \eta) (\partial_\theta \beta \epsilon) - \frac{(\partial_\theta \epsilon)(q)}{\dr 1(q)} (\partial_k q) \dr 1(\theta) + \sum_{j} \partial_k \gamma_j \frac{\partial (\beta \epsilon)}{\partial \gamma_j}.
\end{eqnarray}
Non-symmetric Fermi weights occur whenever odd Lagrange parameters are non-zero and the low temperature approximation is slightly modified to 
\begin{eqnarray}
\label{eq:sommerfeldGeneral_nonsym}
    \int_{-\infty}^\infty d\theta \, n(\theta) f(\theta)
    &=&\int_{q_L}^{q_R} d\theta \, f(\theta) + \frac{\pi^2 T^2}{6} \left[ \frac{f'(q_R) - \left(\frac{ \epsilon''(q_R)}{\epsilon'(q_R)} - U(q_R)\right) f(q_R)}{(\epsilon'(q_R))^2} -\frac{f'(q_L) - \left(\frac{ \epsilon''(q_L)}{\epsilon'(q_L)} - U(q_L)\right) f(q_L)}{(\epsilon'(q_L))^2}\right] , \nn\\
&=& \int_{q_L}^{q_R} d\theta \, f(\theta) + \Lambda_f(q_R) - \Lambda_f(q_L).
\end{eqnarray}
Based on the definition of the density and $\eta$ fields, then at low temperatures the left right Fermi points are considered as $q_{L/R}(T) = \pm q + \delta_{L/R}(T) $. At zero temperature the Fermi functions $\delta_{L/R}(T) \overset{T\to 0}{=}0$ and so
\begin{eqnarray}
  \Lambda_\rho(\theta) &=& \frac{\pi^2 T^2 (\dr 1(\theta))}{6 (\epsilon'(\theta))^2} \left( 2 V(\theta) (\dr 1(\theta)) + \left[ \frac{\epsilon''(\theta)}{\epsilon'(\theta)} - U(\theta) \right] (\dr 1(\theta))\right) \nn\\
    \rho &=& \int d\theta \, n \dr 1  \approx \int^{q_R}_{q_L} d\theta \, \dr 1 -   \Lambda_\rho(q_R) +   \Lambda_\rho(q_L) \approx \int^{q_R}_{q_L} d\theta \, \dr 1 -   \Lambda_\rho(q) +   \Lambda_\rho(-q) + O(T^2 \delta(T)), \nn\\
    &=& \rho_0 + \dr 1(q) (\delta_R + \delta_L) -   \Lambda_\rho(q) +   \Lambda_\rho(-q) \nn\\
    0 &=& \int d\theta \, n \dr \theta - \rho \eta \sim \int^{q_R}_{q_L} d\theta \, \dr \theta +O(T^2 \delta(T))\nn\\
    &=& \frac{q_R + q_L}{2} 2(\dr 1(q')) \dr \theta(q')+ \dr \theta (q) (\delta_R + \delta_L) .
\end{eqnarray}
Taken together these expressions confirm that the corrections to the Fermi points are $O(T^2)$ in the low temperature limit.

\subsection{General viscosity expression eq.~\eqref{eq:generalViscosity}}
Based on the work done so far diffusion for general conserved charges, denoted $q_n = \theta^n/n!$, can be computed. This is carried out by noting the general form of the diffusion kernel 
\begin{eqnarray}
  D(q_n) &=&\partial_x \int \frac{d\theta d\mu}{2}  \frac{\dr{(\theta^n)}(\theta)}{n!} (\dr 1(\theta)) \tilde D(\theta,\mu) \partial_x n(\mu) .
\end{eqnarray}
The kernel simplifies in the thermal case or more generally in the case for $n(q) = n(-q)$ find
\begin{eqnarray}
\frac{\partial_x (\beta \epsilon (\mu))}{\partial_\mu \epsilon} v^{\rm eff}(\mu)-    \frac{\partial_x( \beta \epsilon (\theta))}{\partial_\theta \epsilon} v^{\rm eff}(\theta) =\beta \left( v^{\rm eff}(\theta) - v^{\rm eff}(\mu) \right) \partial_x \eta + \left( \frac{\epsilon(\mu)}{\dr 1(\mu)} - \frac{\epsilon(\theta)}{\dr 1(\theta)} \right)\partial_x \beta , 
\end{eqnarray}
When there is a single Fermi sea the state energy $\beta \epsilon$ a general differential, for $k = x, t$ reads
\begin{eqnarray}
\partial_x(\beta \epsilon) = -(\partial_x \eta) (\partial_\theta \beta \epsilon) - \frac{(\partial_\theta \beta \epsilon)(q)}{\dr 1(q)} (\partial_x q) \dr 1(\theta) + \sum_{j} \partial_x \gamma_j \frac{\partial (\beta \epsilon)}{\partial \gamma_j}.
\end{eqnarray}
this expression can be conveniently rewritten with the anti-symmetric function  
\begin{eqnarray}
    h(\theta,\mu) =  2\pi (\dr 1(\theta)\dr 1(\mu)) (\dr \varphi(\theta,\mu))^2  \frac{(v^{\rm eff}(\theta) - v^{\rm eff}(\mu))^3}{|v^{\rm eff}(\theta) - v^{\rm eff}(\mu)|}.
\end{eqnarray}
So that at low-temperatures the diffusion for an arbitrary conserved charge has a general form given by 
\begin{eqnarray}
  D(q_n) &=&\partial_x \int \frac{d\theta d\mu}{4}  \left(\frac{\dr{(\theta^n)}(\theta)}{n!(\dr 1(\theta))} - \frac{\dr{(\theta^n)}(\mu)}{n! (\dr 1(\mu))}  \right) \frac{\partial_\theta n}{\partial_\theta \epsilon} \frac{\partial_\mu n}{\beta\partial_\mu \epsilon}  h(\theta,\mu) \partial_x \eta \sim O(T).
\end{eqnarray}
As in the case of the energy diffusion the conserved charge $q_n$ can be boosted into the co-moving frame by $\theta \to \theta + \eta$. Such a calculation involves expansions of $\dr{\left[ (\theta + \eta)^n - (\alpha + \eta)^n\right]}/n!$, which are then simplified by noting $\dr{(\theta^n)}(q) = (-1)^n \dr{(\theta^n)}(-q)$. None of this will impact the leading temperature order correction, which is always $O(T)$.

In addition to these general considerations there is a correction to the spatial derivative of the thermal energy, which reads
\begin{eqnarray}
\label{eq:newDefin_lagrange}
    \frac{\partial_x \epsilon }{\partial_\theta \epsilon} = - \partial_x \eta + \sum_j \frac{\partial_x \gamma_j}{\partial_\theta \epsilon(\theta)} \frac{\partial  \epsilon}{\partial \gamma_j} \mp  \frac{v^{\rm eff}(\pm q)}{v^{\rm eff}(\theta)} \partial_x q = - \partial_x \eta \mp \frac{v^{\rm eff}(\pm q)}{v^{\rm eff}(\theta)} \partial_x q + \sum_j \frac{ v_j(\theta)}{v^{\rm eff}( \theta)} \partial_x \gamma_j
\end{eqnarray}
with the additional caveat that $\frac{\partial \epsilon}{\partial \gamma_j } = \dr{(\theta^j)}/ j!$.
With $v_j = \frac{\dr{(\theta^j)}}{j! \, \dr 1(\theta)} $ and the sum of velocities given by $\frac{ \partial_\theta \epsilon}{\dr 1}$ as well as $v_1 = v_F$. A consequence of this term is that the diffusion constant
\begin{eqnarray}
    D^{\alpha}(Q_n) &=&\partial_x \int \frac{d\theta d\alpha}{2}  \dr q_n(\theta) \frac{\partial_\theta n}{\partial_\theta \epsilon} \frac{\partial_\alpha n}{\beta\partial_\alpha \epsilon}   \dr 1 (\alpha) ( \dr \varphi(\theta,\alpha))^2 |v_F(\theta) - v_F(\alpha)| \left[ - \frac {\partial_x (\beta \epsilon(\theta))}{\partial_\theta \epsilon} v^{\rm eff}(\theta) + \frac{ \partial_x (\beta \epsilon(\alpha))}{\partial_\alpha \epsilon} v^{\rm eff}(\alpha) \right].\nn\\
\end{eqnarray}
Insert the ratio for the derivative to find that in the general case
\begin{eqnarray}
    D^{\alpha}(Q_n) &=& \partial_x \int \frac{d\theta d\alpha}{4}  \left(v_n(\theta)-v_n(\alpha)  \right)\frac{\partial_\theta n}{\partial_\theta \epsilon} \frac{\partial_\alpha n}{\beta\partial_\alpha \epsilon}  [\dr 1 (\alpha) \dr 1(\theta)] | v_F(\theta) - v_F(\alpha)| ( \dr \varphi(\theta,\alpha))^2 \nn\\
    &&\left[ (v^{\rm eff}(\theta) - v^{\rm eff}(\alpha))\partial_x \eta - \sum_{j=3} (v_j(\theta) - v_j(\alpha)) \partial_x \gamma_j \right] \nn\\
\end{eqnarray}
For even dressed charges it is straightforward to see that $v_n(q) - v_n(-q) = 0$, whereas $v_n(q) - v_n(-q) = 2 v_n(q)$ for odd values of $n$. An interchange $q \leftrightarrow -q $ leaves the expression unchanged and so the diffusion is non-zero only for odd $n$ and given by
\begin{eqnarray}
    D_n 
    &=& \partial_x \left[ \frac{4 (\dr \varphi(q,-q))^2 v_n(q) v_F(q)}{ \beta v^{\rm eff}(q) v^{\rm eff}(-q)}  \left( v_F(q) \partial_x \eta + \sum_{j = 3, 5, \dots}  \left( v_{j}(q) \gamma_{j+1} \partial_x \eta - v_{j}(q) \partial_x \gamma_j \right)\right) \right] .
\end{eqnarray}
Note that the when the additional Lagrange parameters $\gamma_j=0$ for all $j=3,4 \dots $ the diffusion reduces to the thermal case.

\section{Final result: low-temperature viscous CHD}
\label{app:viscous_ghd_final}
By restoring the conventional hydrodynamic fields with a factor $1/2\pi$ such as $\rho = \frac {1}{2\pi} \int d\theta \, n \dr 1$, and $p = \frac {1}{2\pi} \int d\theta \, \dr \theta n $, (analogously for pressure, energy, and energy), together with 
\begin{equation}
    K = (1^{\rm dr})^2(q)/(2 \pi) = \pi \rho/v_F,  \quad \quad  \partial_\rho v_F =  \frac{1 - 4 \varphi^{\rm dr}(q,-q) v_F}{{K}/\pi } , 
\end{equation}
with the identity $v_F (\dr 1)^2 (q)/\pi = \frac{\dr \theta (q)}{\dr 1 (q)} (\dr 1)^2(q)/\pi = \dr \theta (q) \dr 1 (q)/ \pi = \rho$, 
we finally find
\begin{align}
    \partial_t &  \rho + \partial_x (\eta \rho) = 0,\nn\\
    \partial_t & \eta + \frac{1}{2} \partial_x ( \eta^2) +\frac{\partial_x  \mathcal{P}_s(\rho,T) }{\rho} =  
\frac{1}{\rho} \partial_x (\mu (\rho,T) \partial_x \eta ) ,\nn \\
 ( \partial_t &  T + \eta \partial_x T ) =   - T   \frac{\pi \left( \frac{1}{\rho} + \frac{\partial_\rho v_F}{ v_F}  \right) }{ 3 v_F \tilde{\chi}_e}  \partial_x \eta + \frac{\mu(\rho,T)}{\rho T  \tilde \chi^T_e } (\partial_x \eta)^2,
\end{align}
together with
\begin{eqnarray}
\tilde{\chi}_e &=&  \frac{1}{ {\rho}}  \frac{ \pi }{3 v_F } ,\nn\\
\mu(\rho,T) &=&  \frac{ T }{4 } \frac{ \left(1- \frac{2 K}{2\pi}\partial_\rho v_F\right)^2}{v_F} = 4 T  \dr \varphi(q,-q)^2 v_F ,\nn\\
   \partial_x \mathcal P_s(\rho,T) &=&\Big(  v_F^2 - \frac{\pi T^2}{6} \frac{1- 4 v_F \varphi^{\rm dr}(q,-q)}{\rho v_F} \Big) \partial_x \rho + \frac{ \pi T}{3 v_F} \partial_x T.
\end{eqnarray}

\section{Proofs and checks of various Identities}
\label{app:alltherest}

\subsection{Numerical check of viscosity}
\label{app:numericalCheck_visc}

\begin{figure}[!h]
    \centering
    \includegraphics[width=0.6\columnwidth]{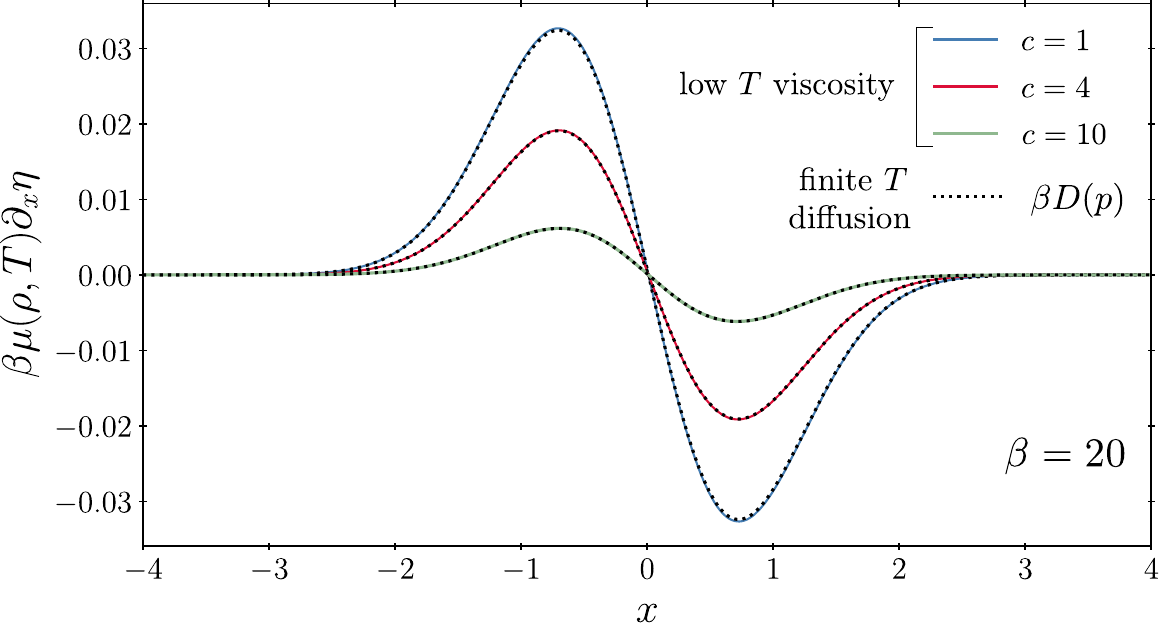}
    \caption{Comparison between the low $T$ viscosity $\mu(\rho,T)\de_x \eta$ appearing in the rhs of~\eqref{eq:viscousGHD} (solid lines) and the finite $T$ diffusion $D(p)$ (dotted lines), see eq.~\eqref{eq:app_diff_kernel} for its expression, where they are found to be an extremely close match. These calculations were performed at $\beta =20$ at various values of $c$. }
    \label{fig:diff_kernel_vs_prediction}
\end{figure}

To verify that our low-temperature result for the viscosity
\begin{eqnarray}
\label{eq:app_visc_final}
    \mu(\rho,T) &=& \frac{ \pi T }{2} \frac{ (1- 2 K\partial_\rho v_F)^2}{v_F} = 8 \pi T \dr \varphi(q,-q)^2 v_F
\end{eqnarray}
is consistent with low-temperature GHD a numerical evaluation of the full diffusion kernel, see eq.~\eqref{eq:app_diff_kernel}, is carried out. The computed diffusion kernel is then specialized to the momentum component and compared to our final result for the low-temperature viscosity. 

Our protocol is as follows: two Gaussian fields are introduced, the potential $\mu(x) = 1 + \mu_0 \e^{-\nu x^2}$ and shifts in the rapidity given by $\eta(x) = \eta_0 \e^{- \sigma x^2}$. These two fields are used to determine the filling function by use of the thermal energy
\begin{eqnarray}
\beta \epsilon(\theta,x) = \beta\frac{(\theta - \eta(x))^2}{2} - \mu(x) - \int d\alpha \, \varphi(\theta,\alpha) \ln \left| 1 + \e^{-\beta \epsilon(\alpha,x)}\right|.
\end{eqnarray}
Once the filling function is determined the full diffusion kernel is considered and applied to the bare charge to determine the general diffusion. The filling function is also used to determine the hydrodynamic density $\rho$ and the Fermi points $\pm q$, which are then used to determine the Fermi velocity $v_F=v^{\rm eff}(q)$ and dressed kernel $\dr \varphi(q,-q)$ thus determining our proposed viscosity. Comparing both methods leads to identical results, see fig.~\ref{fig:diff_kernel_vs_prediction}.  


\subsection{Hydrodynamic fields at $T=0$ and their expressions}
\label{app:HDfieldsT0}
To bridge the gap between generalized and conventional hydrodynamics it's useful to consider the low-temperature values of the hydrodynamic fields. These identities can be evaluated in a fairly straightforward fashion by integrating by parts and thereafter applying the relevant dressed function identities. The hydrodynamic variables are collected below for convenient reference
\begin{eqnarray}
        \rho &=& \int d\theta \, \dr 1 n  =  2 \dr \theta(q) \dr 1(q),\nn\\
    p  &=&\int d\theta \, v \dr 1 n  = \frac 1 2 \left[ \dr{(\theta^2)}(\theta^+)-\dr{(\theta^2)}(\theta^-) \right] \dr 1(q)= \eta \rho ,\nn\\
       \mathcal{ P} &=& \int d\theta \, \dr \theta n \theta = \int d\theta \, \dr \theta_s n \theta_s + \eta^2 \rho = \mathcal P_s + \eta^2 \rho,\nn\\
   E &=& \int d\theta \, \theta^2 \dr 1 n = \frac{1}{6} \left[ \dr{(\theta^3)}(\theta^+)-\dr{(\theta^3)}(\theta^-) \right] \dr 1(q), \nn\\
   E+ \mathcal{ P} &=& \frac 1 2 \left[ [\dr{(\theta^2)} \dr \theta](\theta^+)-[\dr{(\theta^2)} \dr \theta](\theta^-) \right],\nn\\
   E+\mathcal{ P} - \eta^2 \rho &=& \frac 1 2 \left[ \dr{(\theta^2)}(\theta^+)+\dr{(\theta^2)}(\theta^-) \right] \dr \theta(q).
\end{eqnarray}

To extract the hydrodynamic equations from the truncated GHD equations both the space and time derivatives are considered. These relations are exact at zero temperature, however, their subleading temperature correction is $O(\beta^{-2})$. At lowest order in temperature the space/ time derivatives of the hydrodynamic fields can be evaluated explicitly in terms of either hydrodynamic fields or equivalently TBA quantities respectively. These derivative identities, with $i = x,t$, are collected below
\begin{eqnarray}
\partial_i \rho &=& 2( \dr 1(\theta^+))^2(\partial_i q),\nn\\
\partial_i p &=& \rho \partial_i \eta + \eta \partial_i \rho, \nn\\
&=& 2 (\dr 1(\theta^+))^2 \left( v_F\partial_i \eta + \eta (\partial_i q)\right),\nn \\
   \partial_i \mathcal P &=& \partial_i (\rho \eta^2) +v_F^2 \partial_i \rho, \nn\\
   &=& 2 (\dr 1(q))^2 \left[2 (\partial_i \eta)  \eta v_F +  \left( \eta^2 + v_F^2\right) \partial_i q\right], \nn \\
    \partial_i E       &=&\eta \rho \partial_i \eta + \frac{\partial_i q}{v_F} (\mathcal P+E-\eta^2 \rho), \nn\\
    &=& (\partial_i \eta) 2 \eta \dr \theta_s(q) \dr 1 + (\partial_i q)  \left[ \dr{(\theta^2_s)}(q) + \eta^2 \dr 1(q)\right] \dr 1, \nn \\
    \label{eq:zero_T_hydro_deriv}
    \partial_i J_E &=& (\mathcal P+E) \partial_i \eta + \eta v \rho \partial_i q - \eta^2 \rho \partial_i \eta + \eta \partial_i E, \nn\\    &=&\partial_i \eta \left([\dr{(\theta^2_s)}(q) + \eta^2 \dr 1(q)]  \dr \theta(q)  + 2 \eta^2 \dr \theta_s(q) \dr 1(q) \right) + \partial_i q \left( \eta \dr 1(q) [\dr{(\theta^2_s)}(q) + \eta^2 \dr 1(q)]  + 2 \eta( \dr \theta_s(q))^2\right). \nn\\
\end{eqnarray}

\subsection{Proof of the low-temperature expansion of eq. \eqref{eq:niceId}}
\label{app:niceId_deriv}
The goal of this section is to prove the identity
\begin{eqnarray}
    \beta (2 n- 1) \partial_\theta n \to -\sum_{\sigma} \frac{\sigma}{\partial_\theta \epsilon(\theta^\sigma)} \delta'(\theta -\theta^{\sigma}).
\end{eqnarray}
Consider the following integral with a smooth function $f(\theta)$. Expansion of the arguments will allow us to apply well known integral approximation methods to evaluate the result
\begin{eqnarray}
    \int d\theta f(\theta) \beta (2n(\theta) -1) \partial_\theta n = -\int d\theta \, f(\theta) \beta^2 \frac{(1-\e^{\beta \epsilon(\theta)} )\e^{\beta \epsilon} \epsilon'(\theta)}{(1+\e^{\beta \epsilon(\theta)})^3},
\end{eqnarray}
It is key that there are Fermi points that can be identified, specifically $\epsilon(\theta^\pm) =0$, such that expanding around these points the thermal energy may be rewritten as
\begin{eqnarray}
    \beta \epsilon(\theta) &\sim& \beta \epsilon(\theta^\pm) + \beta \epsilon'(\theta^\pm) (\theta -\theta^{\pm)} + \frac \beta 2 \epsilon''(\theta^\pm)(\theta -\theta^\pm)^2 + O((\theta-\theta^\pm)^2),\nn\\
&=& 0 + s + \frac{\epsilon''(\theta^\pm)}{2 \beta (\epsilon'(\theta^\pm))^2 } s^2 + O\left(\frac {1} {\beta^4}\right), \nn\\
\epsilon'(\theta) &\sim& \epsilon'(\theta^\pm) + \epsilon''(\theta^\pm) (\theta - \theta^\pm) + O((\theta -\theta^\pm)^2).
\end{eqnarray}
Transformation of the integration measure to our new measure $s$ is then given by
\begin{eqnarray}
    s = \beta \epsilon'(\theta^\pm) (\theta - \theta^\pm) && d s = \beta \epsilon'(\theta^\pm) d \theta.
\end{eqnarray}
So that the integral under consideration is now given, in terms of these variables $s$. For concreteness consider the upper half of the integral, such that $\theta^+$ is expanded around. Furthermore, introduce $\Lambda = \epsilon''(\theta^+)/(2 (\epsilon'(\theta^+))^2)$ and note that $\theta^+ > \theta^-$ then the integral can be approximated as
\begin{eqnarray}
&    -&\int_{(\theta^+ + \theta^-)/2}^\infty d\theta \, f(\theta) \beta^2 \epsilon'(\theta) \frac{(1-\e^{\beta \epsilon(\theta)} )\e^{\beta \epsilon}}{(1+\e^{\beta \epsilon(\theta)})^3} \sim -\int^\infty_{\beta \epsilon(\theta^+) (\theta^- - \theta^+)/2} d s  \, f\left(\frac{s}{\beta \epsilon'(\theta^\pm)} + \theta^+\right)   \partial_s\left( \frac{\e^{s + s^2 \frac{\Lambda}{\beta}}}{(1+\e^{s + s^2 \frac{\Lambda}{\beta}})^2}\right), \nn\\
&&\sim -\beta\int^\infty_{-\infty} d s  \, \left( f\left(\theta^+\right)+\frac{s}{\beta \epsilon'(\theta^\pm)} f'\left( \theta^+\right) \right)   \partial_s\left( \frac{\e^{s }}{(1+\e^{s} )^2} +\frac{\Lambda}{\beta} \frac{\e^{s } (1- \e^{s }) s^2 }{(1+\e^{s })^3} + O\left( \frac 1 {\beta^2} \right)\right) .
\end{eqnarray}
Then to leading order as $\beta \to \infty$ the integral yields
\begin{eqnarray}
   \int^\infty_{\frac{\theta^+ + \theta^-}{2}} d\theta f(\theta) \beta (2n(\theta) -1) \partial_\theta n &=& - f(\theta^+) \int^\infty_{-\infty} ds\, \beta  \partial_s \frac{\e^{s }}{(1+\e^{s })^2}  - \frac{f'(\theta^+)}{  \epsilon'(\theta^+)} \int^\infty_{-\infty}ds \, s \partial_s \frac{ \e^s}{(1+\e^s)^2} \nn\\
   &&+ f(\theta^+) \Lambda \epsilon'(\theta^+) \int_{-\infty}^\infty ds \, \frac{\e^{s } (1- \e^{s }) s^2 }{(1+\e^{s })^3}, \nn\\
&=& f'(\theta^+) \frac{1}{\epsilon'(\theta^+)}.
\end{eqnarray}
The first term vanishes on integration, likewise the third term is also vanishing after integration by parts leaving only the second term. The procedure is identical with the left Fermi points, albeit with an additional sign flip and so it follows that
\begin{eqnarray}
    \int^\infty_{-\infty} d\theta f(\theta) \beta (2n(\theta) -1) \partial_\theta n = \left( f'(\theta^+) \frac{1}{\epsilon'(\theta^+)} - f'(\theta^-) \frac{1}{\epsilon'(\theta^-)}\right) + O\left( \frac 1 \beta\right).
\end{eqnarray}
Thus identify the final result with the identity~\eqref{eq:niceId}
\begin{eqnarray}
    \beta (2 n -1) \partial_\theta n = -  \sum_\sigma \frac{\sigma}{\partial_\theta \epsilon(\theta^\sigma)} \delta'(\theta - \theta^\sigma).
\end{eqnarray}

\subsection{Specific heat at low temperatures}
\label{app:constantDensitySusc}
The low-temperature expansion for the specific heat $\tilde \chi_e$ can be computed by standard methods
\begin{equation}
 \tilde{\chi}_e =  \frac{1}{\rho} \lim_{T \to 0}T^{-1} \left( \frac{d E (\rho,T)}{d T} \right) \bigg|_{\rho={\rm const}}.
\end{equation}
This will be obtained by determining the low-temperature energy at constant density. Such a quantity is obtained by expanding first around a constant chemical potential, then introducing a changing chemical potential correction that keeps $\rho$ constant. By use of the general formula for a conserved charge~\eqref{eq:general_conservedCharge} at constant chemical potential the low-temperature energy is given in terms of $U(q) = \dr \varphi(q,-q) + \dr \varphi(q,q)$ and $V(q) = \dr \varphi(q,q)-\dr \varphi(q,-q) $, in the comoving frame, as

\begin{align}
    \int d\theta n (\theta)\left(\frac{\theta^2}{2}\right)^{\rm dr} 
    &= \int_{-q(0)}^{q(0)} \dr{\left[ \frac{\theta^2}{2}\right]}_{T=0} +  \frac{\pi^2 T^2}{3(\epsilon'(q))^2} \left(  \frac{(\dr \theta)^2}{v_F} - 2 (\dr 1) V \dr{\left[ \frac{\theta^2}{2}\right]} - 2 (\dr 1)\left(\frac{ 1}{v_F} - 2 U\right) \dr{\left[ \frac{\theta^2}{2}\right]}\right) (q).
\end{align}
To keep density constant the zero temperature Fermi points $q(0)$ must be changed, such that the chemical potential is allowed to vary. Consider a constant density $\rho_{\rm ref} = \rho(q_B)$ and a second low-temperature density at constant chemical potential $\rho(q(T))$. The low-temperature corrections to density are known, see App.~\ref{app:subleading_temperature}, to be 
\begin{eqnarray}
    \rho(q(T)) = \rho(q(0)) - T^2 f(q) = \rho(q(0)) - \frac{\pi^2 T^2}{3  \dr \theta v_F} \left( 2  V(q) (\dr 1(q)) + \left[ \frac {1}{v_F} - 2U(q)\right]  (\dr 1(q)) \right).
\end{eqnarray}
By imposing $\rho(q(T)) = \rho(q_B)$ the condition of constant density requires that
\begin{eqnarray}
    \rho(q(T)) = \rho(q(0)) - T^2 f(q) = \rho(q(0)) + 2 (\dr 1(q(0))) (q_B - q(0)).
\end{eqnarray}
Consequently the Fermi points must satisfy 
\begin{eqnarray}
\label{eq:fermi_point_shift}
    2 (q_B - q(0)) = - \frac{\pi^2 T^2}{3 \dr \theta v_F } \left( 2  V(q) + \left[ \frac 1 {v_F} - 2U(q)\right]   \right),
\end{eqnarray}
to keep the density constant. Now inserting $q(0) = q_B + T^2 f / 2$ the low-temperature correction to energy with constant density is determined to be
\begin{align}
  \rho e(\rho,T) & = \rho e(\rho,0)+  \frac{\pi^2 T^2}{3 \dr \theta v_F } \left( 2  V(q) + \left[ \frac 1 {v_F} - 2U(q)\right]   \right) \dr{\left[ \frac{\theta^2}{2}\right]} ,\nn\\
  &+\frac{\pi^2 T^2}{3(\epsilon'(q))^2} \left(  \frac{(\dr \theta)^2}{v_F} - 2 (\dr 1) V \dr{\left[ \frac{\theta^2}{2}\right]} - 2 (\dr 1)\left(\frac{ 1}{v_F} - 2 U\right) \dr{\left[ \frac{\theta^2}{2}\right]}\right) (q),\nn\\
  &= \rho e(\rho,0) + \frac{\pi^2 T^2 }{3 v_F}.
\end{align}
An application of the temperature derivative on the above formula then implies the sought-after expression for the specific heat
\begin{align}
\tilde \chi_e = \frac{1}{T} \, (\partial_T e)_{\rho = {\rm const}} =\frac{1}{T} \, \partial_T \left( \frac{E_s}{\rho} \right)_{\rho = {\rm const}} = \frac{ 2 \pi^2}{3 \rho v_F }.
  \end{align}

\subsection{Hydrodynamic field corrections at $O(T^2)$}
\label{app:subleading_temperature}

In general a Sommerfeld approximation can be applied to obtain a general low-temperature expansion~\cite{bertini_2018_lowT_transport} 
\begin{eqnarray}
\label{eq:sommerfeldGeneral}
    \int_{-\infty}^\infty n(\theta) f(\theta) 
    &=&\int_{-q}^{q} f(\theta) + \frac{\pi^2 T^2}{6(\epsilon'(q))^2} \left[ f'(q) - f'(-q) - \left(\frac{ \epsilon''(q)}{\epsilon'(q)} - U(q)\right) (f(q) + f(-q))\right].
\end{eqnarray}
This identity is sufficient to determine the subleading temperature corrections to the integrals, which are key to determining the hydrodynamic fields. This expansion as written characterizes the low-temperature expansion for constant chemical potential. 
To account for the low-temperature expansion of the relevant charges it will be relevant to consider both the temperature correction to the dressed function, occurring due to the dressing, and the corrections in the integral noted above. The low-temperature corrections due to dressing a function is considered first. It should also be noted that whether the function being dressed is even or odd also impacts the expansion and so these cases are considered separately.

The low-temperature corrections to a general dressed function $\dr g$ is considered. The dressed functions are assumed to be either symmetric $\dr g(q) = \dr g(-q)$ or anti-symmetric $\dr g(q) = -\dr g(-q)$. Recall the definition for a dressed function
\begin{eqnarray}
    \dr{g}(\theta) = g(\theta) + \int_{-\infty}^\infty \varphi(\theta-\mu) n(\mu) \dr{g}(\mu).
\end{eqnarray}
For the purposes of keeping track of the small temperature limits the zero temperature dressed function is also introduced
\begin{eqnarray}
    \dr g_0(\theta) = g(\theta) + \int^{\infty}_{-\infty} \varphi(\theta-\alpha) n_0(\alpha) \dr g_0(\alpha)= g(\theta) + \int^{q}_{-q} \varphi(\theta-\alpha) \dr g_0(\alpha).
\end{eqnarray}

So the goal is to consider the low T expansion of the integral on the RHS of the dressing relation. So identify the variable for use of the Sommerfeld expansion of eq.~\eqref{eq:sommerfeldGeneral} as
    \begin{eqnarray}
    f(\theta,\alpha) &=& \varphi(\theta-\alpha)\dr {g}(\alpha).\nn\\
    \end{eqnarray}

For even functions, $g(q) = g(-q)$, the low-temperature expansion is carried out with
    \begin{eqnarray}
    f'(\theta,q) - f'(\theta,-q) &=& -[(\partial_\theta \varphi(\theta-q) - \partial_\theta \varphi(\theta+q)) + (\varphi(\theta,q) + \varphi(\theta,-q)) (\dr \varphi(q,q) - \dr \varphi(q,-q))] \dr {g}(q) \nn\\
    &&+ [\varphi(\theta-q) + \varphi(\theta + q)] \dr{g}(q), \nn\\
    f(\theta,q) + f(\theta,-q) &=& (\varphi(\theta,q) + \varphi(\theta,-q)) \dr {g}(q).
\end{eqnarray}
From the above the small temperature correction reads
\begin{eqnarray}
\label{eq:lowT_symm_g}
(    \dr {g} - \dr {g}_0) &=& \frac{ \pi^2 T^2  \dr {g}(q)}{6 (\epsilon'(q))^2} \left(- [\dr{(\partial_\theta \varphi(\theta-q) - \partial_\theta \varphi(\theta+q))}\frac{}{} \right.  \nn\\
&&\left.+ (\dr \varphi(\theta,q) + \dr \varphi(\theta,-q)) (\dr \varphi(q,q) - \dr \varphi(q,-q))] -  \left[\frac{1}{v_F} - 2U(q) \right] (\dr \varphi(\theta,q) + \dr \varphi(\theta,-q)) \right) \nn\\
&&+ \frac{ \pi^2 T^2 \dr{(\partial g)}(q) }{6 (\epsilon'(q))^2} \left[ \dr \varphi(\theta,q) + \dr \varphi(\theta,-q) \right].
\end{eqnarray}
Note that on the RHS all functions are evaluated at zero temperature, however either the zero temperature or finite temperature functions can be used without altering the $O(T^2)$ order and so the subscript is suppressed. 

Now consider the antisymmetric case $g(q)= - g(-q)$. Now the identities for the low-temperature expansion read
   \begin{eqnarray}
    f'(\theta,q) - f'(\theta,-q) &=& -(\varphi'(\theta - q) + \varphi'(\theta+q)) \dr \theta(q) + (\partial_\theta \dr \theta)(q) (\varphi(\theta - q) - \varphi(\theta + q) ),\nn\\
    f(\theta,q) + f(\theta,-q) &=& (\varphi(\theta,q) - \varphi(\theta,-q)) \dr {g}(q).
\end{eqnarray}
With the temperature corrections to $\dr g$ given by
\begin{eqnarray}
(    \dr {g} - \dr {g}_0) &=& \frac{ \pi^2 T^2  \dr {g}(q)}{6 (\epsilon'(q))^2} \left(- [\dr{(\partial_\theta \varphi(\theta-q) + \partial_\theta \varphi(\theta+q))}\frac{}{} \right.  \nn\\
&&\left.+ (\dr \varphi(\theta,q) - \dr \varphi(\theta,-q)) (\dr \varphi(q,q) + \dr \varphi(q,-q))] -  \left[\frac{1}{v_F} - 2U(q) \right] (\dr \varphi(\theta,q) - \dr \varphi(\theta,-q)) \right) \nn\\
&&+ \frac{ \pi^2 T^2 \dr{(\partial g)}(q) }{6 (\epsilon'(q))^2} \left[ \dr \varphi(\theta,q) - \dr \varphi(\theta,-q) \right].
\end{eqnarray}

These relations are then combined with
\begin{eqnarray}
    \int d\theta \, n \dr{[\partial_\theta \varphi(\theta,q)]} &=&         -\int d\theta \, n \dr{[\partial_\theta \varphi(\theta,-q)]} =\dr 1(q) [\dr \varphi(q,q) - \dr \varphi(q,-q)] = \dr 1(q) V(q),\nn\\
    \int d\theta \, n(\theta) \dr \varphi(\theta,q) &=& (\dr 1(q) - 1) ,\nn\\
    \int d\theta \, \theta n \dr{\varphi(\theta,\pm q)} &=&  (\dr \theta - \theta)(\pm q),\nn\\
    \int d\theta \, \theta n \dr{[\partial_\theta \varphi(\theta,\pm q)]} &=&  (1 - \dr 1) +  \dr \theta(q)  U(q).
\end{eqnarray}
As a consequence the even functions are given by
\begin{eqnarray}
\label{eq:lowTevenDress}
    \int d\theta \, n (\dr g-\dr g_0) &=& \frac{\pi^2 T^2 \dr g(q)}{3 (\dr \theta)^2} \left( - 2 (\dr 1) V(q) + V(q) - \left( \frac 1 {v_F} - 2 U(q)+f(\alpha)\right)  (\dr 1 -1)  \right) + \frac{\pi^2 T^2 \dr{(\partial g)}}{3 (\dr \theta)^2} (\dr 1 - 1) ,\nn\\
    \int d\theta \, \theta n (\dr g-\dr g_0) &=& 0.
\end{eqnarray}
And the odd functions satisfy the similar relations
\begin{eqnarray}
\label{eq:lowToddDress}
    \int d\theta \, n (\dr g-\dr g_0) &=& 0 ,\nn\\
    \int d\theta \, \theta n (\dr g-\dr g_0) &=& \frac{\pi^2 T^2 \dr g(q)}{3 (\dr \theta)^2} \left( (\dr 1 - 1)  - (2 \dr \theta(q) - q) U(q) - \left( \frac 1 {v_F} - 2U(q) +f(\alpha) \right) (\dr \theta(q) - q) \right) \nn\\
    &&+ \frac{\pi^2 T^2 \dr {(\partial g)}(q)}{3 (\dr \theta)^2} (\dr \theta(q) - q) .\nn\\
\end{eqnarray}
Noting that a subscript $'0$' denotes that the zero temperature value of a function is being considered. With the functions $f(\alpha)$ introduced as potential corrections due to additional Lagrange parameters in the thermal energy, which are expected to take the form
\begin{eqnarray}
    \frac{\epsilon''}{\epsilon'} \sim \frac{1}{v_F(\theta)} - U(\theta) +f(\alpha)
\end{eqnarray}

\subsubsection*{General expression for conserved charges and currents at constant $\mu(q)$ and low $T$}
\label{app:general_lowTcorr}
A general hydrodynamic charge field is given by $g(\theta) = \theta^n / n!$ as
\begin{eqnarray}
    Q_n = \int d\theta\, \dr{g}(\theta) n(\theta).
\end{eqnarray}
Such the Sommerfeld expansion~\eqref{eq:sommerfeldGeneral} can be applied by identifying
\begin{eqnarray}
    f(\theta) &=& \dr g,\nn\\
    f'(q) - f'(-q) &=& 2\dr{(\partial g)}(q) - 2 \left( \dr \varphi(q,q) - (-1)^n \dr \varphi(q,-q)   \right) \dr g(q).
\end{eqnarray}
When the function $g(\theta)$ is antisymmetric the low-temperature corrections to the corresponding conserved charge vanish. As a consequence consider only the case where $g(\theta)$ is symmetric
\begin{eqnarray}
    Q_n(T) = 
    \int n_0 \dr{g} + \frac{\pi^2 T^2}{3 (\dr \theta)^2} \left( \dr{(\partial g)} - \dr g(q) V(q) - \left( \frac 1{v_F} -  2 U(q) \right) \dr g(q)\right).
\end{eqnarray}
This must be combined with the low $T$ expansion for the even dressed functions of eq.~\eqref{eq:lowTevenDress} and simplified. The overall result for the low-temperature even conserved charges is found to be
\begin{eqnarray}
\label{eq:general_conservedCharge}
    Q_n(T) - Q_n(0) 
= \frac{\pi^2 T^2 \dr 1(q) }{3 (\epsilon'(q))^2} \left[\left( 2 (U(q) -V(q))  - \frac{ 1}{v_F} \right) \dr g(q) + \dr{(\partial g)}(q) \right].
\end{eqnarray}
In addition to the general conserved charge, the general conserved current can be written out. Specifically
\begin{eqnarray}
   \mathcal J_n = \int d\theta \, v g \dr 1 n = \int d\theta \, \theta \dr g n.
\end{eqnarray}
So identify the following to insert into the approximation formula
\begin{eqnarray}
    f(\theta) &=& \theta \dr g ,\nn\\
    f'(q) - f'(-q) &=& 2 \dr g(q) + 2 q\dr{(\partial g)}(q) - 2 q U(q) \dr g(q) .
\end{eqnarray}
Unlike the care of the hydrodynamic charges in the case of the currents only antisymmetric $g(\theta)=\theta^n/n!$ need be considered. From the general expression we then find
\begin{eqnarray}
    \mathcal J_n(T) = \int d\theta \, \theta n_0 \dr g + \frac{\pi^2 T^2}{3 (\dr \theta)^2} \left( \dr g(q) + q \dr{(\partial g)}(q) -  q U(q) \dr g(q) - \left( \frac{1}{v_F} - 2 U(q)\right) q \dr g(q) \right).
\end{eqnarray}
This is combined with the identity for temperature corrections to the odd dressed functions, eq.~\eqref{eq:lowToddDress}. With the final result being
\begin{eqnarray}
   \mathcal J_n(T)- \mathcal J_n(0) 
    &=&   \frac{\pi^2 T^2 \dr {(\partial g)} \dr \theta(q)}{3 (\epsilon'(q))^2} .
\end{eqnarray}

These can be used to incorporate more complicated forms of the thermal energy by restoring the $f(\alpha)$ factor to find the charge and current corrections at constant chemical potential as
\begin{eqnarray}
        Q_n(T) - Q_n(0)&=& \frac{\pi^2 T^2 \dr 1(q) }{3 (\epsilon'(q))^2} \left[\left( 2 (U(q) -V(q))  - \frac{ 1}{v_F} -f(\alpha) \right) \dr g(q) + \dr{(\partial g)}(q) \right],\nn\\
  \mathcal J_n(T)- \mathcal J_n(0) 
    &=&   \frac{\pi^2 T^2 \dr {(\partial g)} \dr \theta(q)}{3 (\epsilon'(q))^2} .
\end{eqnarray}
It should also be noted that these describe the behaviour of a symmetric Fermi sea, which can be generalized by shifting the both Fermi seas separately in the multi-Fermi sea scenario.

\subsubsection*{Low-temperature pressure at constant density}
\label{app:pressure_corr}
Temperature corrections to the pressure $\mathcal P_s = \mathcal{J}_1$ can be determined from these equations, noting that 
\begin{eqnarray}
    \mathcal P_s = \int_{-\infty}^\infty d\theta \,\dr \theta  \theta n .
\end{eqnarray}
So that the result at low-temperature with constant chemical potential is  
\begin{eqnarray}
     \mathcal{P}_s -\mathcal P_s(0)
&=& \frac{\pi^2 T^2}{3 v_F}.
\end{eqnarray}
Likewise, the density at finite temperature is determined by the integral
\begin{eqnarray}
    \rho = \int d\theta \, \dr 1(\theta) n(\theta).
\end{eqnarray}
By application of the even conserved charge identity the temperature correction to density is given by
\begin{eqnarray}
    \rho - \rho(0) 
    &=&- \frac{\pi^2 T^2}{3 v_F \dr \theta} \left( 2  V(q) (\dr 1(q)) + \left[ \frac 1 {v_F} - 2U(q)\right]  (\dr 1(q)) \right).
\end{eqnarray}
With these results for the density and pressure at fixed $q$ consider the pressure at a constant low-temperature reference state $\rho_{\rm ref}$. For the density to be constant it must be the case that
\begin{eqnarray}
 \rho_{\rm ref} &=&  \rho(q(T)) = \rho(q_B) , \nn\\
&=& \rho(q(0)) - T^2 f(q(0)) =  \rho(q(0)) + \frac{\delta \rho}{\delta q} \bigg|_{q = q(0)}(q_B -q(0)) .
\end{eqnarray}
With $f(q(0))$ being the small temperature correction to density at constant chemical potential $\mu(q)$. As a consequence the temperature correction for the pressure in a fixed state $\rho_{\rm ref}$ is given by
\begin{eqnarray}
    \mathcal{P}(\rho(q(T))) &-& \mathcal{P}(\rho(q_B)) = \mathcal{P}(\rho(q)) + \frac{\pi^2 T^2}{3 v_F} - \mathcal{P}(\rho(q)) - \frac{ \delta \mathcal{P}(\rho(q))}{\delta q} \bigg|_{q = q(0)}(q_B - q(0)) , \nn\\
    &=&  \frac{2\pi^2 T^2}{3  } \left(   \frac 1 {v_F} +V(q) - U(q) \right),\nn\\
     &=&  \frac{2\pi^2 T^2}{3 v_F } \left(   1 - 2 {v_F} \dr \varphi(q,-q) \right) 
     =  \frac{\pi^2 T^2}{3 v_F } ( 1 + \partial_q v_F) = \frac{\pi^2 T^2}{3 v_F} ( 1 +  2 K \partial_\rho v_F),
\end{eqnarray}
with the note that $\partial_\rho \mathcal P = v_F^2$ .

\subsection{Dressed functions identities at low temperature}
\label{app:id_list}
For considering the spatial derivative of the thermal energy it is possible to construct a relation
\begin{eqnarray}
    \label{eq:ratio_id}
\frac{\partial_x \epsilon }{\partial_\theta \epsilon} = -\partial_x \eta - \frac{v_F}{v_s(\theta)} \partial_x q,
\end{eqnarray}
the utility of which is that the spatial derivative of a thermal sea with arbitrary Fermi points is put into correspondence with variables relying only on symmetric Fermi points.

At zero temperature the derivative of the dressed functions can be explicitly written out in terms of other dressed functions. These identities are collected here. These identities appear frequently at various intermediate steps.
\begin{eqnarray}
    \partial_\theta \dr f \bigg|_{\theta = \pm q}&=& \dr{(\partial_\theta f)} - \dr \varphi(\pm q,q) \dr f(q) + \dr \varphi(\pm q,-q) \dr f(-q) ,\nn\\
    (\partial_q \dr f) \bigg|_{\theta =\pm q} &=& \dr \varphi(\pm q,q) \dr f(q) + \dr \varphi(\pm q,-q) \dr f(-q) ,\nn\\
    \label{eq:zero_temp_fluct}
    \partial_q (\dr f(q)) &=& \dr {(\partial f)}(q) + 2 \dr \varphi(q,-q) \dr f(-q).
\end{eqnarray}
With these relations many identities can be written out for other composite quantities like the effective velocity
\begin{eqnarray}
    \partial_\theta v_s(\theta) \bigg|_{\theta=q} &=& 1 - 2 v_F \dr \varphi(q,-q),\nn\\
    \partial_q v_s(\theta) \bigg|_{\theta= q} &=&  - 2 v_F \dr \varphi(q,-q) ,\nn\\
    \label{eq:velocity_id}
    \partial_q (v_s(q))&=& 1 - 4 v_F \dr \varphi(q,-q),
\end{eqnarray}
as well as $v_E = \dr{(\theta^2)}/(2 (\dr 1))$, where
\begin{eqnarray}
\partial_x (v_E(\theta) )\bigg|_{\theta = q} 
&=& - v_F \partial_x \eta ,\nn\\
\partial_\theta v_E(\theta) \bigg|_{\theta = q} 
&=& v_F.
\end{eqnarray}
Finally, the dressed kernel derivative identities are given by
\begin{eqnarray}
    \partial_x \dr \varphi(q,-q)   &=& 2 (\partial_x q) \dr \varphi(q,q) \dr \varphi(q,-q),\nn \\
    \partial_q \dr \varphi(\theta,\mu) \bigg|_{\theta,\mu= q,-q}   &=& 2  \dr \varphi(q,q) \dr \varphi(q,-q).
\end{eqnarray}
Consistent with the more general identities previously noted.\\

\subsection{Diffusion at large coupling }
\label{app:tg_diff_deriv}
In the large $c$ limit, keeping the leading term, the diffusion can be more directly computed. Many of the steps are similar to the interacting scenario, however due to explicit form of the thermal energy, 
\begin{equation}
    \beta \epsilon = \beta \frac{(\theta - \theta^-)(\theta - \theta^+)}{2} + O\left( \frac 1 c \right),
\end{equation}
it is possible to write out the relations in a more explicit form. This section is included for completeness, however the results are completely consistent with the $c \to \infty$ limit keeping only the $O(1/c^2)$ terms. In the large $c$ limit $\dr \varphi \sim 1/(\pi c) $ so that the leading contribution to the diffusion kernel yields
\begin{eqnarray}
  \frac 1 2 \partial_x( \int d\alpha  [  \tilde D_{  \infty}] \partial_x n) 
&\overset{c \gg 1}{\sim}&    \frac 1 {2\pi c^2}  \int d\alpha  |\theta- \alpha|(\theta - \alpha) \partial_x \left[\frac{\partial_\theta n(\theta)}{\partial_\theta \epsilon}  \frac{\partial_\alpha n(\alpha)}{\beta  \partial_\alpha \epsilon}  ( \partial_x \theta^+ + \partial_x \theta^-) \right].
\end{eqnarray}
This may be more easily simplified by recalling that
\begin{eqnarray}
    \frac{\partial_\theta n}{\partial_\theta \epsilon} = - \sum_{\sigma = \pm} \sigma \frac{ \delta(\theta - \theta^\sigma)}{\theta - \frac{(\theta^+ + \theta^-)}{2}}.
\end{eqnarray}
So find that the integral becomes
\begin{eqnarray}
    \frac 1 2 \partial_x \left( \int d\alpha  [  \tilde D_{  \infty}] \partial_x n\right)  &\overset{c \gg 1}{\sim}& -   \frac 1 {2\pi c^2} \partial_x \left( \sum_{\sigma= \pm} \sigma  |\theta- \theta^\sigma|\frac{(\theta - \theta^\sigma)}{\theta^\sigma - \eta}   \frac{\partial_\theta n(\theta)}{\beta \partial_\theta \epsilon}    ( \partial_x \theta^+ + \partial_x \theta^-) \right).
\end{eqnarray}
Note that the presence of the $\delta$-function under the $\partial_x$ complicates the picture, since now acting on the integral introduces a $\delta'$-function that cannot be matched by other terms. This represents the fluctuations around the edge of the Fermi seas. Explicitly,
the low-temperature diffusion kernel in the large $c$ limit is given by
\begin{eqnarray}
\label{eq:app_TG_diff_fp}
\frac 1 2 \partial_x\left( \int d\alpha [\tilde D_\infty] \partial_x n\right)&=&\frac 4 {\pi c^2\beta} \sum_\sigma \sigma \delta(\theta - \theta^\sigma) \partial_x^2 \eta+   \frac{4}{\pi c^2} \sum_{\sigma,\sigma'} \left( \frac{|\theta -\theta^{\sigma'}|(\theta -\theta^{\sigma'}) }{\beta (\theta^+ - \theta^-)^2}\right) \delta'(\theta -\theta^\sigma) \frac{\partial_x( \beta \epsilon(\theta))}{\beta \partial_\theta \epsilon} (\partial_x \eta),\nn\\
&& -\frac{8}{\pi c^2} \sum_\sigma \delta(\theta-\theta^\sigma) \frac{(\partial_x \eta)^2}{\beta(\theta^+ - \theta^-)} - \frac{4 q}{\pi c^2 \beta^2} (\partial_x \eta)(\partial_x \beta) \sum_ \sigma \delta(\theta - \theta^\sigma).
\end{eqnarray}
This an be combined with with Euler GHD in terms of a single evolution equation for the Fermi weights
\begin{align}
    &\partial_t n + v \partial_x n = \frac 1 2 \partial_x\left( \int d\alpha [\tilde D_\infty] \partial_x n\right) , \nn\\
   \sum_\sigma \sigma \delta(\theta - \theta^\sigma)&\left( \partial_t \theta^\sigma + v^{\rm eff}(\theta) \partial_x \theta^\sigma \right)  = \sum_ \sigma \sigma \delta(\theta - \theta^\sigma) \left(\frac 4 {\pi c^2\beta}  \partial_x^2 \eta -\frac{8 \sigma}{\pi c^2}  \frac{(\partial_x \eta)^2}{\beta(\theta^+ - \theta^-)} - \frac{4 q \sigma}{\pi c^2 \beta^2} (\partial_x \eta)(\partial_x \beta) \right)\nn\\
   &+   \frac{4}{\pi c^2} \sum_{\sigma,\sigma'} \left( \frac{|\theta -\theta^{\sigma'}|(\theta -\theta^{\sigma'}) }{\beta (\theta^+ - \theta^-)^2}\right) \delta'(\theta -\theta^\sigma) \left(\frac{\partial_x \beta}{\beta} \frac{\epsilon(\theta)}{\partial_\theta \epsilon} - \partial_x \eta - \frac{q}{\partial_\theta \epsilon} \partial_x q \right)(\partial_x \eta).
\end{align}
In the second line the term containing $\partial_x \beta$ should be noted as vanishing, since $\epsilon(\theta^\sigma) = 0 $. Consequently, by inserting a test function and simplifying it follows that the $\partial_x \beta$ term from the first and second lines are either identically zero or exactly cancel.  Subsequent removal of the test function then implies
\begin{align}
   \sum_\sigma \sigma \delta(\theta - \theta^\sigma)&\left( \partial_t \theta^\sigma + v^{\rm eff}(\theta) \partial_x \theta^\sigma \right)  = \sum_ \sigma \sigma \delta(\theta - \theta^\sigma) \left(\frac 4 {\pi c^2\beta}  \partial_x^2 \eta -\frac{8 \sigma}{\pi c^2}  \frac{(\partial_x \eta)^2}{\beta(\theta^+ - \theta^-)}  \right)\nn\\
   &-  \frac{4}{\pi c^2} \sum_{\sigma,\sigma'} \left( \frac{|\theta -\theta^{\sigma'}|(\theta -\theta^{\sigma'}) }{\beta (\theta^+ - \theta^-)^2}\right) \delta'(\theta -\theta^\sigma) \left(  \partial_x \eta + \frac{q}{\partial_\theta \epsilon} \partial_x q \right)(\partial_x \eta).
\end{align}

Now the $\delta'$-function must be dealt with. By considering the diffusive corrections to conserved charges, which allows us to make sense of the $\delta'$-function. When integrating over $\int d\theta \, 1$ the result is a vanishing correction. A less trivial result is obtained by considering the integration of $\int d\theta \, \theta$, and $\int d\theta \, \theta^2/2$, corresponding to the momentum and energy diffusion respectively
\begin{eqnarray}
\frac 1 2 \int d\theta \, \theta \partial_x\left( \int d\alpha [\tilde D_\infty] \partial_x n\right)&=&\frac {8 } {\pi c^2 \beta} \left(q \partial_x^2 \eta+   (\partial_x q)(\partial_x \eta)- \frac{\partial_x \beta}{\beta} q \partial_x \eta\right)  = \frac{8}{\pi c^2} \partial_x \left( \frac{q \partial_x \eta}{\beta} \right), \nn \\
\label{eq:free_diffusion_app}
\frac 1 2 \int d\theta \, \frac{ \theta^2}{2} \partial_x\left( \int d\alpha [\tilde D_\infty] \partial_x n\right)&=&\frac {8 \eta} {\pi c^2 \beta} \left(q \partial_x^2 \eta+   (\partial_x q)(\partial_x \eta)- \frac{\partial_x \beta}{\beta} q \partial_x \eta\right)+ \frac {8 q} {\pi c^2 \beta} \left(\partial_x \eta\right)^2.
\end{eqnarray}
Note that this implies the viscosity in the large $c$ limit, after identification of $\rho = 2 q$, as being given by
\begin{eqnarray}
    \mu(\rho,T) = \frac{4 T}{\pi c^2} \rho ,
\end{eqnarray}
this result is consistent with the $c \gg 1$ limit of the low-temperature diffusion with interaction. \\


\subsection{Commutativity of low-temperature limit and spatial derivative}
\label{app:commutativity}

Along with the above, the $T=0$ limit commutes with the spatial derivative. That this commutes is demonstrated below by considering the zero temperature Fermi weight in the large $c$ limit and an LDA of the Fermi weight with
\begin{eqnarray}
    n(\theta) = \Theta(\theta -\theta^-) - \Theta(\theta - \theta^+) = -\sum_{\sigma = \pm} \sigma \Theta(\theta - \theta^\sigma).
\end{eqnarray}
The spatial derivatives are then computed as
\begin{eqnarray}
    \partial_x n(\theta) &=& \sum_{\sigma= \pm } \sigma \delta(\theta - \theta^\sigma) \partial_x \theta^\sigma, \nn\\
    \partial_x^2 n(\theta) &=& \sum_{\sigma = \pm} \left( \sigma \delta (\theta - \theta^\sigma) (\partial_x^2 \theta^\sigma) -\sigma \delta' (\theta - \theta^\sigma) (\partial_x \theta^\sigma)^2 \right) .
\end{eqnarray}
This result will now be contrasted with the finite temperature case. Application of the derivatives in this case makes use of the Eq.~\eqref{eq:niceId}, $\eta = \frac{\theta^+ + \theta^-}{2}$, and $q = \frac{\theta^+ - \theta^-}{2}$ to write the finite temperature second spatial deriative as
\begin{eqnarray}
    \partial^2_x n &=&  - \sum_{\sigma = \pm} \sigma \delta (\theta - \theta^\sigma)  \frac{\partial_x^2 \epsilon}{\partial_\theta \epsilon}  - \sum_{\sigma = \pm} \frac{\sigma}{\partial_\theta \epsilon(\theta^\sigma)} \delta' (\theta - \theta^\sigma) \frac{(\partial_x \epsilon)^2}{\partial_\theta \epsilon}, \nn\\
    &=&- \sum_{\sigma = \pm} \sigma \delta (\theta - \theta^\sigma)  \frac{[(\theta^- - \theta)\partial_x^2 \theta^+ + (\theta^+ - \theta)\partial_x^2 \theta^-] +  2\partial_x \theta^+ \partial_x\theta^-}{2 (\theta - \eta)} \nn\\
    &&- \sum_{\sigma = \pm} \frac{1}{q} \delta' (\theta - \theta^\sigma) \frac{[(\theta - \theta^{-})\partial_x \theta^++(\theta - \theta^{+})\partial_x \theta^-]^2}{4(\theta - \eta)}.
\end{eqnarray}
For clarity consider the $\delta'(\theta-\theta^\sigma)$ term on its own. Interpretation of the $\delta'(\theta - \theta^\sigma)$ motivates the application of an overall integral over $\theta$. So inserting the expressions for the derivatives of the energy $\epsilon$ have
\begin{eqnarray}
    &&\int d\theta  \sum_{\sigma = \pm} \frac{1}{q} \delta' (\theta - \theta^\sigma) \frac{[(\theta - \theta^{-})\frac{\partial_x \theta^+}{2}+(\theta - \theta^{+})\frac{\partial_x \theta^-}{2}]^2}{4(\theta - \eta)} f(\theta) ] \nn \\
    &&= - \left[ (\partial_x \theta^++\partial_x \theta^-) \sum_\sigma \partial_x( \theta^\sigma) \frac{ f(\theta^\sigma)}{q} - \sum_\sigma (  \partial_x \theta^\sigma )^2 \frac{f(\theta^\sigma)}{q}+ [(\partial_x \theta^+)^2-(\partial_x \theta^-)^2] \partial_\theta f(\theta)\right],\nn\\
&&= -  \sum_\sigma  \delta(\theta -\theta^\sigma) \left[  \sigma (\partial_x \theta^\sigma)^2(\partial_\theta  f)(\theta) +\frac{  (\partial_x \theta^+) (\partial_x \theta^-)}{q} f(\theta) \right].
\end{eqnarray}
This is now recombined with the $\delta(\theta -\theta^\sigma)$ term and the equivalence between the two expressions follows.

\end{document}